\documentclass[fleqn,usenatbib]{aa} %

\makeatletter
\let\do@linenumbers\relax
\let\linenumbers\relax

\makeatother

\usepackage[varg]{txfonts}

\DeclareRobustCommand{\VAN}[3]{#2}
\let\VANthebibliography\thebibliography
\def\thebibliography{\DeclareRobustCommand{\VAN}[3]{##3}\VANthebibliography}

\usepackage{graphicx}	
\usepackage{amsmath}	
\usepackage[version=4]{mhchem}     

\usepackage{mathtools}

\usepackage{hyperref}
\hypersetup{colorlinks=true,linkcolor=blue,citecolor=blue,filecolor=blue,urlcolor=blue, breaklinks=True}

\usepackage{booktabs}   

\usepackage[referable,para,flushleft]{threeparttablex} 

\newcommand{\kepler}{\textit{Kepler}\xspace}

\newcommand{\numax}{\ensuremath{\nu_{\mathrm{max}}}\xspace}

\newcommand{\msol}{\ensuremath{\mathrm{M}_\odot}\xspace}

\makeatletter
\newcommand\thefontsize[1]{{#1 The current font size is: \f@size pt\par}}
\makeatother

\usepackage{subcaption}

\begin{document}

\title{Parametric models of core-helium-burning stars: structural glitches near the core}

\author{Massimiliano Matteuzzi\inst{\ref{inst1},\ref{inst2}}\thanks{E-mail: \href{mailto:massimilia.matteuzz2@unibo.it}{massimilia.matteuzz2@unibo.it}}
\and
Gaël Buldgen\inst{\ref{inst3}}
\and
Marc-Antoine Dupret\inst{\ref{inst3}}
\and \\
Andrea Miglio\inst{\ref{inst1},\ref{inst2}}
\and 
Lucy Panier\inst{\ref{inst3}}
\and
Walter E. van Rossem\inst{\ref{inst1}}
}

\institute{Department of Physics \& Astronomy "Augusto Righi", University of Bologna, via Gobetti 93/2, 40129 Bologna, Italy\label{inst1}
\and
INAF-Astrophysics and Space Science Observatory of Bologna, via Gobetti 93/3, 40129 Bologna, Italy\label{inst2}
\and
Space Sciences, Technologies and Astrophysics Research (STAR) Institute, Université de Liège, Allée du Six-Août 19, 4000 Liège, Belgium\label{inst3}
}
\abstract{
Understanding the internal structure of core helium burning (CHeB) stars is essential for evaluating transport processes in regions where nuclear reactions occur, developing accurate models of stellar populations, and assessing nucleosynthesis processes that impact the chemical evolution of galaxies. Until recently, detailed insights into the innermost layers of these stars were limited. However, advancements in asteroseismic observations have allowed us to explore their stratification more thoroughly. Despite this progress, the seismic signatures associated with structural variations at the boundary between the convective and radiative core, as well as the chemical composition gradients within the radiative core, have been relatively underexplored in CHeB stars. This paper aims to fill that gap by investigating how these gradients influence the oscillation modes of low-mass CHeB stars. We specifically focus on mixed dipole modes and uncoupled g-modes as effective probes of stellar interiors.
Using semi-analytical models calibrated with the evolutionary codes \texttt{BaSTI-IAC}, \texttt{CLES}, and \texttt{MESA}, we explore the influence of density discontinuities and their associated structural glitches on the period spacings of these oscillation modes.
These codes were chosen for their distinct physical prescriptions, allowing us to identify common relevant features for calibration purposes.
This approach enables us to control the type of glitch introduced while maintaining a realistic representation of the star.
As expected from previous studies, our results indicate that these glitches manifest as distinct periodic components in the period spacings, which can be used to infer the position and amplitude of the structural variations inside CHeB stars.
Furthermore, we compare models with smooth transitions to those with sharp discontinuities, highlighting the differences in period spacing and mode trapping, which permits us to infer the sharpness of the glitches. Additionally, we conduct simulations based on four-year-long \kepler observations, demonstrating that our models predict oscillation frequencies closely resembling the observed data. Ultimately, our models enable realistic predictions of how each sharp structural variation impacts the observed power spectral density. This alignment not only validates our theoretical approach but also suggests promising directions for interpreting glitches signatures in high-quality asteroseismic data.
}

\keywords{
Asteroseismology -- Stars: evolution -- Stars: fundamental parameters -- Stars: horizontal-branch -- Stars: interiors -- Stars: low-mass
}


\titlerunning{Parametric models of core-helium-burning stars: structural glitches near the core}
\authorrunning{Matteuzzi et al.}
\maketitle

\section{Introduction}
\label{sec:intro}
Understanding the extent of mixing within the helium core is crucial to refine estimates of core helium burning (CHeB) phase lifetimes and improve predictions of stellar internal structures in subsequent evolutionary stages. This knowledge plays a key role in developing accurate models of stellar populations and elucidating the chemical evolution of galaxies. Observations of mixed dipole modes in CHeB stars provide a valuable tool for probing their innermost regions \citep[e.g.][]{Montalban2010,Bedding2011,Mosser2012c,Mosser2012b,Montalban2013,Deheuvels2016,Vrard2016,Mosser2017,Deheuvels2018,Mosser2018,Mosser2024}. These oscillation modes may reveal structural variations that occur at the boundaries of mixed convective regions, in areas of element ionisation, or at interfaces between layers with different chemical compositions resulting from nuclear burning. In particular, we expect that CHeB stars exhibit chemical composition gradients at the boundary between convective and radiative core, within the hydrogen-burning shell, and at the base of the convective envelope \citep[e.g.][]{Ledoux1947,Schwarzschild1958,Castellani1971a,Kippenhahn2012}. In low-mass stars with a degenerate helium core, the transition from the red giant branch (RGB) to the CHeB phase is expected to involve a succession of off-centre helium flashes \citep[e.g.][]{Kippenhahn2012}, which induce additional chemical composition gradients within the radiative core.
\begin{figure}[htbp]
\centering
\includegraphics[width=\columnwidth]{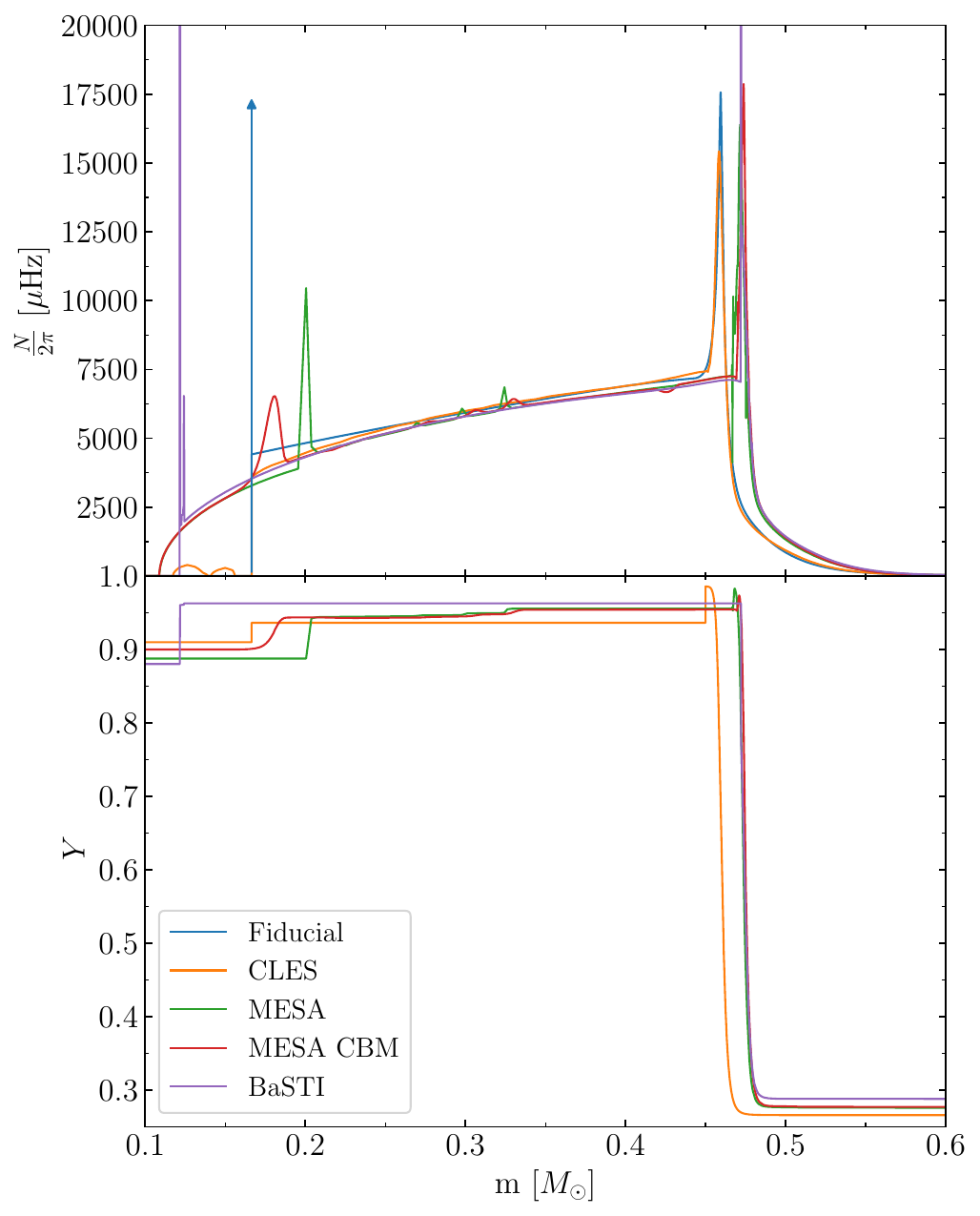} 
\caption{Brunt-Väisälä frequencies as a function of internal mass for five different star models at the beginning of the CHeB stage (top panel). The orange, green, red, and violet lines represent models calculated using the \texttt{CLES}, \texttt{MESA}, and \texttt{BaSTI-IAC} evolutionary codes (see Section \ref{sec:fidmod}), while the blue line corresponds to the fiducial barotropic model discussed in Section \ref{sec:fidmod}. All models are for a 1 \msol CHeB star with solar metallicity. The red line represents a \texttt{MESA} model computed with the boundary mixing prescriptions outlined in Appendix \ref{sec:boundary_layer_MESA}. These models reveal common features in their Brunt-Väisälä frequencies, including a convective core, radiative core, hydrogen-burning shell, radiative envelope, and convective envelope (the latter not shown for clarity). The bottom panel shows the helium mass fraction as a function of internal mass for four of the five models.}
\label{fig:fiducial_model}
\end{figure}

In particular, as shown in Figure \ref{fig:fiducial_model}, the temperature ($\nabla := d \ln T/ d\ln P$), adiabatic [$\nabla_\mathrm{ad} := \left( \partial \ln T/ \partial \ln P \right)_S$] and mean molecular weight ($\nabla_\mu := d \ln \mu / d\ln P$) gradients manifest as signatures in the squared Brunt-Väisälä frequency 
\begin{equation}
\label{eq:Brunt_def}
\begin{aligned}
    N^2 (r) & := \frac{G m(r)}{r^2} \left[ \frac{1}{\Gamma_1} \frac{d \ln P(r)}{d r} - \frac{d \ln \rho(r)}{d r} \right]  \\ &
    \approx \frac{G^2 m^2 \rho}{r^4 P} \left( \nabla_\mathrm{ad} - \nabla + \nabla_\mu  \right) ,
\end{aligned}
\end{equation}
which determines the behaviour of oscillation modes. This frequency can also produce deviations in period spacing \citep[$\Delta P$, defined in e.g.][]{Aerts2010} from the asymptotic value when the characteristic scales of structural variations within the core are comparable to or smaller than the local wavelength of the waves under investigation \citep[see][for a review]{Cunha2020}. These distinct signatures on the eigenfrequencies are commonly referred to as glitch signatures. 

Previous studies have examined near-core mixing conditions through the asymptotic period spacing of dipole modes $\Delta P_\mathrm{a}$ \citep[e.g.][]{Bossini2015,Constantino2015,Bossini2017,Noll2024}, but further insights can be gained by analysing individual eigenfrequencies. Despite extensive observational and theoretical studies on these glitches \citep[e.g.][]{Miglio2008,Bossini2015,Cunha2015,Mosser2015,Cunha2019,Vrard2022,Hatta2023,Cunha2024}, the effects of sharp variations in the density profile on eigenfrequencies remain not fully explored. A significant challenge in the detection of glitch signatures arises from the inherent complexity of the spectra, which exhibit a densely packed pattern of mixed modes. This complicates the isolation of individual effects, such as magnetic fields, rotation and strong gradients in chemical composition, and increases the difficulty in interpreting the spectra themselves. Therefore, models are needed to make the interpretation of the observed spectra more reliable.

The seismic signature of density discontinuities associated with abrupt composition changes has already been addressed by \citet{McDermott1990}, who noted that such jumps can significantly influence the pulsation mode spectrum and stability of non-radially oscillating stars. A density discontinuity can locally contribute to or even become the dominant source of buoyancy. Such a discontinuity is defined as any region where the density changes so rapidly that the density scale height becomes much smaller than the effective wavelength of any oscillation mode of interest. The role of these density discontinuities in the spectrum of gravity modes was also explored by \citet{Gabriel1979}, who found that abrupt density changes can lead to mode trapping of g-modes with the proper effective wavelength. If the mode frequency associated with a density discontinuity is comparable to or larger than the ordinary g-mode frequency, these discontinuities cannot be ignored in computing the g-mode spectrum.

Structural glitches give rise to periodic components in the period spacing, allowing for the recovery of information about the location and sharpness of these glitches from their periodicity and amplitude.
Therefore, deviations of the period spacing from $\Delta P_\mathrm{a}$ contain information about sharp features in the $N$ frequency, which can be visualised through the normalised buoyancy radius
\begin{equation}
\label{eq:buoyancy_radius}
\Phi(r) := \frac{ \int_{r_1}^{r} N/y \, dy}{ \int_{r_1}^{r_2} N/y \, d y}, \quad N > 0, \quad r_1 \leq r \leq r_2.
\end{equation}
This quantity is a monotonically increasing function between the inner turning point $r_1$ and the outer turning point $r_2$, effectively serving as a radial coordinate. The coordinate $\Phi(r)$ is particularly valuable for studying the core/envelope symmetry of high-overtone modes, though this symmetry is approximate and can be disrupted by various mechanisms \citep{Montgomery2003}. Within the JWKB approximation \citep[developed by Jeffreys, Wentzel, Kramers and Brillouin, see e.g.][]{Unno1989,Gough2007}, the kinetic energy density per unit of normalised buoyancy radius is generally constant, except at sharp features such as composition transition zones. This property makes variations in kinetic energy density a clear indicator of mode trapping. Moreover, within the JWKB approximation, the period of the component in the period spacing ($P_\mathrm{glitch}$) associated with a structural glitch located at a radius $r_\mathrm{glitch}$ is related to its normalised buoyancy radius [$\Phi(r_\mathrm{glitch})$] and its periodicity in radial order ($\Delta n_\mathrm{glitch}$) via the relation:
\begin{equation}
\label{eq:Tau_signal_Dk}
\Delta n_\mathrm{glitch} \approx 1 / \Phi(r_\mathrm{glitch}) \approx P_\mathrm{glitch} / \Delta P_\mathrm{a}
\end{equation}
\citep[see e.g.][]{Miglio2008,Cunha2015,Cunha2019,Vrard2022,Hatta2023}. This equation is particularly useful for interpreting the periodic signals observed in the period spacing of the oscillation modes, enabling the extraction of information about the location, amplitude, and type of structural glitches.

Consequently, glitch signatures play a significant role in advancing our understanding of the complex case of mixing in CHeB stars, especially in the absence of a robust mixing theory. In this context, we face challenges related to the management of an expanding convective core and the development of semiconvective layers \citep[e.g.][]{Castellani1971b,Straniero2003}. Furthermore, evolutionary models may generate artificial glitch signatures due to numerical inaccuracies in their predictions, complicating the interpretation of the eigenfrequencies. To address these challenges, flexible and self-consistent models are essential, particularly for studying low-mass CHeB stars with high coupling factors \citep[e.g.][]{Matteuzzi2023,Matteuzzi2024}, as they offer deeper insights into their internal structures compared to other red clump (RC) stars.

The paper is organised as follows. Section \ref{sec:model} describes the theoretical framework for solving the differential equations of stellar structure. Section \ref{sec:fidmod} details the semi-analytical model of a 1 \msol RC star with solar composition and $Y_c \approx 0.9$, which serves as a reference model for subsequent modifications. Section \ref{sec:smoothdiscont} explains how we simulate different levels of mixing between adjacent zones in a self-consistent manner. Results are presented in Section \ref{sec:results}, and Section \ref{sec:conc} concludes the paper.

\section{Parametric modelling of CHeB stars based on realistic stellar structures}
\label{sec:model}
The structure of a single, spherical, non-rotating, non-magnetic star can be described by the hydrostatic and mass continuity equations \citep[see e.g.][]{Kippenhahn2012}.
To fully solve these differential equations, it is essential to establish a relation between pressure and density. Furthermore, this relation must include temperature, mass fractions of chemical species, and their corresponding chemical potentials to formulate a complete system of differential equations. In particular, we must incorporate an equation of state (EOS) that connects pressure with these other variables, thereby resulting in a closed system of equations.
In this paper, we will employ a differential representation of an EOS. We start by defining the variable
\begin{equation}
\label{eq:gamma_local}
\gamma_{\rm local}(r) := \frac{d \ln P(r)}{d \ln \rho},
\end{equation}
which allows us to derive the differential equation
\begin{equation}
\label{eq:gamma_loc_diff}
\frac{d P(r)}{dr} = \gamma_{\rm local}(r)  \frac{ P(r) }{\rho(r)} \frac{d \rho(r)}{d r}
\end{equation}
that has to be solved together with the hydrostatic and mass continuity equations. All the information provided by the EOS is encapsulated in equation \ref{eq:gamma_local}, which can be seen through a heuristic approach. For simplicity, and without loss of generality, we may consider $\gamma_{\rm local}(r) \equiv \gamma_{\rm local}(T, \rho, \mu)$. By applying the chain rule to equation \ref{eq:gamma_local}, we can demonstrate that
\begin{equation}
\label{eq:gamma_local_prop}
\gamma_{\rm local}(T, \rho, \mu) = \frac{\chi_{\rho}}{ 1 - \chi_{T} \nabla - \chi_\mu \nabla_\mu } \\
%
,
\end{equation}
where $\chi_\rho := \left. \frac{\partial \ln P}{\partial \ln \rho} \right|_{T,\mu}$, $\chi_\mu := \left. \frac{\partial \ln P}{\partial \ln \mu} \right|_{\rho,T}$, $\chi_T := \left. \frac{\partial \ln P}{\partial \ln T} \right|_{\rho,\mu}$. It is now more evident that $\gamma_{\rm local}$ is predominantly influenced by the thermal and chemical gradients present within the star, and that in adiabatically stratified layers $\gamma_{\rm local} \equiv \Gamma_1$.
Moreover, according to equation \ref{eq:gamma_loc_diff}, a polytropic EOS can be derived (i.e. $P \propto  \rho^{\gamma_{\rm local}} $) when $\gamma_{\rm local}$ is treated as a constant.
For the purposes of the following discussion, we will denote $\gamma_{\rm local}(r) \equiv \gamma(r)$.

\subsection{Barotropic CHeB stars}
\label{sec:barstar}
As demonstrated in Section \ref{sec:model}, a closed set of equations is obtained when we provide the temperature profile and the chemical species. However, stars can be accurately modelled as systems in barotropic equilibrium (i.e. the thermodynamic variables are stratified as the density profile), because $\vec{\nabla} \rho \times \vec{\nabla} P = \vec{0}$ everywhere in self-gravitating spherical stars in hydrostatic equilibrium. Furthermore, the density is a monotonic function of the radius with the exception of a small region near the surface where a density inversion due to hydrogen recombination occurs \citep[e.g.][]{Chitre1967,Ergma1971,Harpaz1984}. To provide a more quantitative estimate of the deviation from monotonicity, we analyse the density inversion region in a representative stellar model using the code \texttt{CLES} (see Section \ref{sec:fidmod}). Our findings indicate that a non-injective region in the function $\rho(P)$ appears only very close to the surface (at $\approx 99.8 \%$ of the total stellar radius) with an extent that constitutes only $0.14 \%$ of the total stellar radius and about $10^{-5} \%$ of the total stellar mass. Therefore, despite the relatively high values of $d P/P \approx 44\%$ and $d \rho / \rho \approx 5\%$ in this region, its very limited size combined with a very low mass content means that it has a negligible effect on the global structure of the star. This observation leads to the conclusion that, with a high degree of accuracy, $P(r) \equiv P(\rho)$, $T(r) \equiv T(\rho)$, $X_i(r) \equiv X_i(\rho)$ and $\gamma(r) \equiv \gamma(\rho)$.
These relations can be incorporated into equation \ref{eq:gamma_loc_diff} to derive the differential equation
\begin{equation}
\label{eq:gamma_loc_diff_barotr}
\frac{d P(r)}{dr} = \frac{ \gamma(\rho) P(\rho) }{\rho(r)} \frac{d \rho(r)}{d r}.
\end{equation}
A general solution of equation \ref{eq:gamma_loc_diff_barotr} is
\begin{equation}
\label{eq:EOS_baro}
P(\rho) = P_c \exp{\left(  \int_{\ln \rho_c}^{\ln \rho} \gamma(t) \, dt \right)},
\end{equation}
where $P_c$ and $\rho_c$ are the central pressure and density, respectively.
Therefore, an a priori knowledge of $\gamma(\rho)$, e.g. from calibrations on evolutionary models, allows us to find the internal structure of the star using the hydrostatic equation, the mass continuity equation, and equation \ref{eq:gamma_loc_diff_barotr} or \ref{eq:EOS_baro}.

\subsubsection{Differential equations for barotropic stars}
\label{sec:diffeq}
From now on we normalise all the quantities respect to reference density ($\rho_1$) and pressure ($P_1$) values. As a result, once the dimensionless quantities 
\begin{equation}
\begin{dcases}
\theta := \frac{\rho}{\rho_1} \\
\xi := r \sqrt{ \frac{4 \pi G\rho_1^2}{P_1} } \\
\beta := \frac{P}{P_1} \\
\psi := m \sqrt{ \frac{4 \pi G^3 \rho_1^4}{P_1^3} }
\end{dcases}
\end{equation} 
are defined, the hydrostatic equation, the mass continuity equation, equation \ref{eq:gamma_loc_diff_barotr} and equation \ref{eq:EOS_baro} become
\begin{equation}
\begin{dcases}
\label{eq:mass_density_diffeq}
\frac{d \theta(\xi)}{d \xi} = - \frac{\psi(\xi) \theta(\xi)^2}{\gamma(\theta) \beta(\theta) \xi^2}\\
\frac{d \psi(\xi)}{d \xi} = \xi^2 \theta(\xi) \\
\beta(\theta) = \beta_c \exp{\left(  \int_{\ln \theta_c}^{\ln \theta} \gamma(t) \, dt \right) },
\end{dcases}
\end{equation}
with initial conditions $\theta(\xi = 0) = \theta_c$, $\beta(\xi = 0) = \beta_c$ and $\psi(\xi = 0) = 0$. We numerically solve equation \ref{eq:mass_density_diffeq} with a Dormand and Prince Runge-Kutta fifth-order method from the centre to the surface and obtain the $\theta(\xi)$ and $\psi(\xi)$ profiles. Finally, with such profiles we a posteriori calculate other variables of interest such as the Brunt-Väisälä frequency
\begin{equation}
N^2(\xi) = 4 \pi G \rho_1 \frac{\psi(\xi)^2 \theta(\xi)}{\beta(\theta)\xi^4} \left[ \frac{1}{\gamma(\theta)} - \frac{1}{\Gamma_1}\right].
\end{equation}
To mitigate numerical issues when solving equation \ref{eq:mass_density_diffeq} from the centre, we employ Taylor expansions to approximate the actual solutions near the centre. More details regarding these expansions are provided in Appendix \ref{app:Taylor_exp}. We also want to notice that in certain situations it may be more beneficial to solve alternative differential equations rather than equation \ref{eq:mass_density_diffeq}. In Appendix \ref{app:num_issues}, we examine some of these alternative equations, with a specific focus on equations near the stellar surface. In Appendix \ref{app:plummer}, we present analytical solutions to equation \ref{eq:mass_density_diffeq} for constant values of $\gamma(r)$, which we utilise to validate our numerical solver.

\subsection{Discontinuities in the internal profiles}
\label{sec:jumpcond}
In CHeB stars we expect that certain internal profiles (e.g. $\rho$) exhibit discontinuities or non-differentiable points. For example, we expect a sharp decrease in density at the boundary between convective and radiative core due to an abrupt change in the chemical profile (see Section \ref{sec:intro}). Therefore, we can establish the jump conditions at a given radial position $r_0$ by ensuring the continuity of pressure and mass. In fact, in CHeB stars at equilibrium, there are no internal shock waves, and the locations where the internal profiles show discontinuities form a countable set with zero Lebesgue's measure. This ensures that the integrals of these internal profiles are continuous. For simplicity let us use the notation $\lim_{r \to r_0^-} f(r) := f^-$ and $\lim_{r \to r_0^+} f(r) := f^+$ for an internal profile $f(r)$. Such a profile is continuous when $f^- = f^+ = f(r_0)$.
Let us define
\begin{equation}
\Lambda := \frac{\rho^+}{\rho^-} \in \mathbb{R^+},
\end{equation}
because we are mainly interested in jump discontinuities in the density profile. Then, the jump conditions are
\begin{equation}
\label{eq:jump_cond}
\begin{dcases}
\xi^+ = \xi^-  \\
\beta^+ = \beta^- \\
\psi^+ = \psi^-  \\
\theta^+ = \theta^- \Lambda  \\
\left.  \frac{d \theta}{d \xi} \right|_{r_0^+} = \left. \frac{d \theta}{d \xi} \right|_{r_0^-} \Lambda^2  \left(  \frac{\gamma^-}{\gamma^+}  \right) \\
\left.  \frac{d \psi}{d \xi} \right|_{r_0^+} =  \left. \frac{d \psi}{d \xi} \right|_{r_0^-}  \Lambda \\
(N^2)^+ =  (N^2)^- \Lambda \left( \frac{\Gamma_1^+ - \gamma^+}{\Gamma_1^- - \gamma^-}  \right)  \left(  \frac{\gamma^-}{\gamma^+}  \right) \left(  \frac{\Gamma_1^-}{\Gamma_1^+}  \right).
\end{dcases}
\end{equation}

Let us now introduce $r_0$ as the boundary between two zones with different normalisation constants $P_1$ and $\rho_1$. This choice may be useful when equation \ref{eq:mass_density_diffeq} creates numerical issues (for example with small $\theta$ and $\beta$ values, see Appendix \ref{app:num_issues}). We then define as normalisation constants for the first zone (i.e. for $r < r_0$ ) the density and the pressure in the centre of the star (i.e. $P_1 \equiv P_c$ and $\rho_1 \equiv \rho_c$). This simplifies all the equations, because now $\theta_c \equiv 1$ and $\beta_c \equiv 1$. In the second zone (i.e. where $r > r_0$) it is useful to define as normalisation constants the pressure and the density at the boundary between the two zones. Therefore, we have $P_1 \equiv P(r_0) = P_c \exp{\left(  \int_{0}^{\ln \theta^-} \gamma(t) \, dt \right) }$ and $\rho_1 \equiv \Lambda \rho^-$. The new jump conditions are now
\begin{equation}
\label{eq:jump_cond_new}
\begin{dcases}
\xi^+ = \xi^- \frac{\Lambda \theta^-}{\sqrt{\beta^-}  } \\
\beta^+ = 1 \\
\psi^+= \psi^- \frac{(\Lambda\theta^-)^2}{(\beta^-)^{\frac{3}{2}} } \\
\theta^+ =1 \\
\left.  \frac{d \theta}{d \xi} \right|_{r_0^+} = \left. \frac{d \theta}{d \xi} \right|_{r_0^-}   \frac{\gamma^-}{\gamma^+} \frac{ \sqrt{ \beta^-   }}{(\theta^-)^2} \\
\phi^+ = 1 \\
\left.  \frac{d \phi}{d \xi} \right|_{r_0^+} = (\gamma^+ -1)  \left. \frac{d \theta}{d \xi} \right|_{r_0^-}  \frac{\gamma^-}{\gamma^+} \frac{\sqrt{ \beta^-}}{(\theta^-)^2},
\end{dcases}
\end{equation}
where we include the $\phi$ and $\frac{d \phi}{d \xi}$ variables defined in Appendix \ref{app:num_issues}, because useful for Section \ref{sec:thirdzone}.

\section{Fiducial barotropic model}
\label{sec:fidmod}
In this section, we derive a fiducial barotropic model for a 1 \msol star with solar composition at the onset of the CHeB stage, ensuring it incorporates all the fundamental characteristics of this evolutionary phase.
As discussed in Section \ref{sec:barstar}, knowing $\gamma(\rho)$ allows us to find the internal structure of the star using the hydrostatic equation, the mass continuity equation, and equation \ref{eq:gamma_loc_diff_barotr} or \ref{eq:EOS_baro}. However, this information is insufficient on its own; we need detailed evolutionary models for the calibration of $\gamma(\rho)$. For this purpose we employ the evolutionary codes \texttt{CLES v21.0} \citep[Code Liégeois d'Évolution Stellaire,][]{Scuflaire2008a} and \texttt{MESA v11701} \citep[Modules for Experiments in Stellar Astrophysics;][]{Paxton2011,Paxton2019}.
In both codes we focused on models that have a mass fraction of helium in the centre of about 0.9 (i.e. $Y_c \approx 0.9$).
In particular, both in \texttt{CLES v21.0} and in \texttt{MESA v11701} we adopt as the reference solar mixture that from \citet{Asplund2009}, and high- and low-temperature radiative opacity tables are computed for the solar specific metal mixture. The envelope convection is described by the mixing length theory of \citet[][]{Cox1968}; the corresponding $\alpha_\mathrm{MLT}$ parameter, the same for all the models, is derived from the solar calibration with the same physics. Below the convective envelope, we add a diffusive undershooting \citep{Herwig2000} with a size parameter $f = 0.02$ \citep[see][]{Khan2018} in the \texttt{MESA v11701} models, while the \texttt{CLES v21.0} models incorporated a step undershooting in the form of penetrative convection. Furthermore, additional mixing over the convective core limit during the CHeB phase is treated following the formalism by \citet{Bossini2017} in the \texttt{MESA v11701} models. In contrast, in the \texttt{CLES v21.0} models we adopt a step overshooting of $0.50 H_P$ in the form of penetrative convection, along with prescriptions designed to establish a semiconvective region where $\nabla_\mathrm{rad} = \nabla_\mathrm{ad}$ \citep{Castellani1971b}. Unlike \texttt{MESA v11701}, \texttt{CLES v21.0} begins its simulations from a zero-age horizontal branch, employing a fixed initial helium core mass of 0.45 \msol. As a result, these models do not undergo helium flashes, but approximately 5\% by mass of helium in the core is converted into carbon (more details in Panier et al., submitted). Lastly, we also simulate a diffusive convective boundary mixing process with \texttt{MESA}, as detailed in Appendix \ref{sec:boundary_layer_MESA}.

\begin{figure}
\centering
\includegraphics[width=\linewidth]{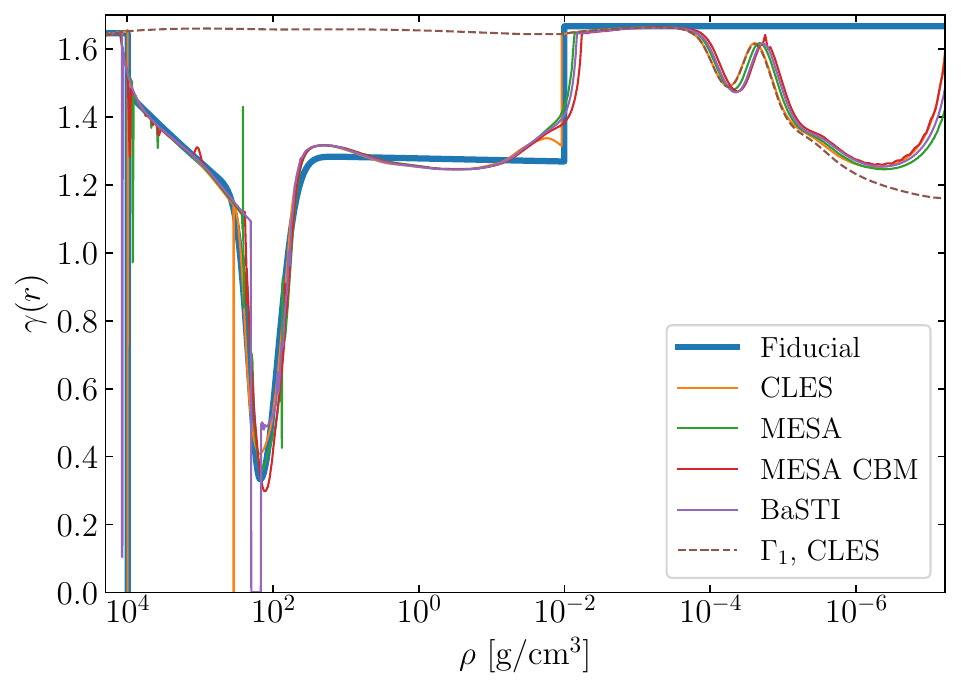} 
\caption{Comparison of $\gamma(r)$ as a function of density for the five different models shown in Figure \ref{fig:fiducial_model}. The dashed brown line represents the $\Gamma_1$ used by \texttt{CLES v21.0} for reference. All five models share the fundamental characteristics of the CHeB phase.}
\label{fig:gamma_all_zones} 
\end{figure}
In Figure \ref{fig:gamma_all_zones} we show $\gamma(r)$ of five different models as a function of density. Three of them are computed with \texttt{CLES v21.0} and \texttt{MESA v11701}, while another one is a model with $Z = 0.012580$ taken from the \texttt{BaSTI-IAC} library\footnote{\url{http://basti-iac.oa-abruzzo.inaf.it/astero.html}} explained in \citet{Hidalgo2018}. It is clear that they share the same fundamental characteristics of the CHeB phase. We expect a similar situation for models of different masses and/or metallicity, provided that they retain equivalent internal structures. Specifically, these models consist of a convective core sorrounded by helium-rich material, an hydrogen-burning shell at the end of the helium core and a convective envelope. The main difference being the location (in density) of such characteristics. Finally, the last $\gamma(r)$ in the figure is a reference barotropic model we derive in Section \ref{sec:firstzone}, \ref{sec:secondzone} and \ref{sec:thirdzone}, which we will call fiducial barotropic model. This model is calibrated on the evolutionary model computed with the code \texttt{CLES v21.0}, which we will simply call reference \texttt{CLES} model. We decide to divide $\gamma(r)$ of the barotropic model in three different zones: the convective core (Section \ref{sec:firstzone}), the inner radiative region of the star (i.e. the radiative core, the hydrogen-burning shell and the radiative envelope, see Section \ref{sec:secondzone}), and the convective envelope (Section \ref{sec:thirdzone}).

It is essential to emphasise that the primary aim of this semi-analytical modelling is to explore the details of the gravity modes spectrum. Therefore, our analysis will concentrate on the characteristics of the stellar radiative regions, while we will simplify the modelling of the convective envelope, as it is less relevant to our investigation.

\subsection{Convective core}
\label{sec:firstzone}
\begin{table}
\centering
\resizebox{\linewidth}{!}{%
\begin{threeparttable}
\centering
\caption{Comparison between the reference \texttt{CLES} model and the fiducial barotropic model presented in Section \ref{sec:fidmod}.}
\label{tab:calib_result}
\begin{tabular}{@{}llllll@{}}
\toprule
Zone  & $\Gamma_1$ & Radius &  Mass & Density & Pressure \\ \midrule
Convective core &  1.6460    &  $< 10^{-10} \% $ & $< 10^{-10}  \%$ & $ 0.08 \%$ & $0.06 \%$ \\ \midrule
Radiative region, correct BCE &  $ 5 /3$   &  $ < 10^{-10} \% $ & $9.3 \cdot 10^{-10}  \%$ & $ 1.3 \cdot 10^{-8} \%$ & $-7.2 \%$ \\ \midrule
Radiative region, fiducial &  $ 5 /3$   &  $< 10^{-10} \% $ & $0.56 \%$ & $ 8.2 \%$ & $7.3 \%$ \\ \midrule
\end{tabular}
\tablefoot{This comparison is done by means of the expression $\left( x_\mathrm{CLES} - x_\mathrm{fiducial} \right) / x_\mathrm{CLES}$, where $x$ is the position, mass, density, and pressure at the boundaries of the zones depending on the column. The second row presents the values achieved when fitting the correct radius and mass at the base of the convective envelope (Section \ref{sec:secondzone}). In contrast, the last row displays the values obtained when aiming for the correct total stellar mass and radius (Section \ref{sec:secondzone}).}
\end{threeparttable}
}
\end{table}
The first zone is the convective core of the CHeB star. We model this part by using the same $\rho_c$ and $P_c$ as the reference \texttt{CLES} model, but a different $\Gamma_1$. We assume a constant $\Gamma_1$ equal to $\gamma(\theta)$, i.e. we assume a polytropic EOS in this region. We calibrate the barotropic model to have the same radius and mass at the end of the convective core as the reference \texttt{CLES} model by keeping $\Gamma_1$ as a free parameter to fit. The results of the calibration are in Table \ref{tab:calib_result}, which clearly indicates that maintaining a fixed $\Gamma_1$ prevents the accurate representation of pressure and density at the end of the convective core.

\subsection{Radiative region}
\label{sec:secondzone}
In the second zone, we simulate the change in chemical composition at the boundary between convective and radiative core with a jump in density identical to the reference \texttt{CLES} model (i.e. we adopt the same $\Lambda<1$). This approach is justified by the fact that $\Lambda \approx \mu^+ / \mu^-$ in the perfect gas approximation, because at the boundary between convective and radiative core the pressure and the temperature are continuous functions. In this second zone, we model the star from the base of the radiative core to the base of the convective envelope. In particular, we take
\begin{equation}
\begin{dcases}
\label{eq:second_zone_case}
\Gamma_1 = \frac{5}{3} \\
0 \leq \gamma(\rho)<\Gamma_1 \\
\theta := \frac{\rho}{\Lambda \rho^-} \\
\beta := \frac{P}{P_0} = \exp{\left(  \int_{0}^{\ln \theta} \gamma(t) \, dt \right) }
\end{dcases}
\end{equation}
with
\begin{equation}
\label{eq:gamma_second_zone}
\gamma(u) =
\begin{dcases}
A \exp{ \left[  - \frac{\left( u - u_P  \right)^2  }{2 \sigma^2}    \right]   }  +  a \exp{\left( K u \right) } , \quad u_P \le u \le 0 \\\\
[A + a \exp{\left( K u_P \right) } - (a') \exp{\left( K' u_P \right) }] \exp{ \left[  - \frac{\left( u - u_P  \right)^2  }{2 (\sigma')^2}    \right]   }  + \\
+ (a') \exp{\left( K' u \right) }, \quad u < u_P  \, \, ,
\end{dcases}
\end{equation}
where we define $u := \ln \theta$ and $u_P := \ln \theta_P$ as useful variables to simplify subsequent calculations.
Therefore, we model $\gamma(\theta)$ in this zone with eight parameters (i.e. $[A, \theta_P, \sigma, a, K, \sigma', K', a']$).
The normalised pressure is found when applying equation \ref{eq:gamma_second_zone} into $\beta(\theta)$ of equation \ref{eq:second_zone_case}, that is
\begin{equation}
\label{eq:ln_beta_second_zone}
\ln \beta(\theta) =
\begin{dcases}
 A \sqrt{\frac{\pi}{2}} \sigma\left[ \mathrm{erf} \left( \frac{\ln \theta - \ln \theta_P}{ \sqrt{2} \sigma} \right)  + \mathrm{erf} \left( \frac{ \ln \theta_P}{ \sqrt{2} \sigma} \right)  \right] + \\
 + \frac{a}{K} \left( \theta^K - 1  \right)  , \quad \theta_P \le \theta \le 1 \\\\
A \sqrt{\frac{\pi}{2}} \sigma\left[ \mathrm{erf} \left( \frac{ \ln \theta_P}{ \sqrt{2} \sigma} \right)  \right]+ \frac{a}{K} \left( \theta_P^K - 1  \right) + \\
+ [A + a \theta_P^K - (a') \theta_P^{K'}] \sqrt{\frac{\pi}{2}} (\sigma')  \mathrm{erf} \left( \frac{\ln \theta - \ln \theta_P}{ \sqrt{2} \sigma'} \right) + \\
+ \frac{a'}{K'} \left( \theta^{K'} -  \theta_P^{K'}  \right), \quad  \theta < \theta_P \, \, .
\end{dcases}
\end{equation}
We find initial guesses of the eight parameters by performing a fit between equation \ref{eq:gamma_second_zone} and the $\gamma(\theta)$ of the reference \texttt{CLES} model with a maximum likelihood estimator. Subsequently, we modify the values of $K'$ and $a'$ to obtain the same radius and mass at the base of the convective envelope (BCE) as in the reference \texttt{CLES} model (see Table \ref{tab:calib_result}). This table further illustrates that the modelling is sufficiently accurate to yield an accurate density for the BCE, despite variations in pressure. However, in the final fiducial model we do not keep these values of $K'$ and $a'$, because they would lead to unphysical values of the total stellar mass and radius (Section \ref{sec:thirdzone}).

Finally, we want to notice that in the reference \texttt{CLES} model there is another drop in density located at the end of the helium core as a signature of the helium-flashes (i.e. the drop in $\gamma$ of Figure \ref{fig:gamma_all_zones} near $\rho = 100$ g/cm$^3$). However, in our fiducial barotropic model we exclude such signature, because we will explore better additional jump discontinuities in Section \ref{sec:results}.

\subsection{Convective envelope}
\label{sec:thirdzone}
In the reference \texttt{CLES} model, the density is non-differentiable at the BCE due to the implemented adiabatic step undershooting explained in Section \ref{sec:fidmod}. This means that $\gamma$ has (with abuse of notation) a "jump discontinuity" at the BCE (see the "jump" in $\gamma$ as illustrated in Figure \ref{fig:gamma_all_zones} at $\rho = 0.01$ g/cm$^3$). This characteristic is retained in our barotropic model.
Moreover, in contrast to the convective core, the convective envelope (designated as the third zone in our barotropic model) exhibits low efficiency in convective transport in the near-surface layers, primarily due to their low density values. As a result, these near-surface layers are characterised by a temperature gradient that exceeds the adiabatic temperature gradient, which can be modelled with the mixing length theory \citep[e.g.][]{Cox1968}. However, in our barotropic model we adopt a $\gamma(\rho) \equiv \Gamma_1$ (i.e. an adiabatic temperature gradient) everywhere in the convective envelope to simplify subsequent calculations. Additionally, we do not account for helium and hydrogen recombination and we keep a constant $\Gamma_1 = 5/3$ (similarly to the second zone, see Section \ref{sec:secondzone}). This simplification implies that the semi-analytical models we compute cannot be used to measure acoustic glitch signatures such as the helium and hydrogen partial ionisation regions. Nevertheless, our models show the expected pattern of p-mode frequencies, characterised by modes that are nearly equally spaced in frequency (see Section \ref{sec:fiducial_model}). Finally, we determine the total stellar mass and radius by solving the Lane-Emden equation \ref{eq:Lane_Emden_final}, with the appropriate jump conditions described in equation \ref{eq:jump_cond_new}, until we reach a density $\rho = 0$.

Due to the simplifications implemented, particularly within the convective envelope, our barotropic model does not yield the same total mass and radius as the reference \texttt{CLES} model; instead, it results in a star that is larger and more massive. A correct total radius or a correct total mass can be achieved when $\Gamma_1 > 5/3$ within the convective envelope. However, we decide to retain a more physically plausible value of $\Gamma_1 = 5/3$ and adjust the parameters $[K', a']$\footnote{We also tested different parameters to reach the correct total mass and radius, but these two are the best compromise.} to ensure that the barotropic model attains the correct total mass and radius. The change in radius, mass, density and pressure at the BCE resulting from the introduction of the new $[K', a']$ parameters are presented in Table \ref{tab:calib_result}.

\section{Smooth transitions in discontinuous density profiles} 
\label{sec:smoothdiscont}
In real stars, both macroscopic (see Appendix \ref{sec:boundary_layer_MESA}) and microscopic \citep[e.g.][]{Michaud1984,Michaud2010} diffusion processes are expected to smooth gradients in pressure, temperature, concentration and density. They tend to smooth out such gradients when present, as in the case of jump discontinuities in density. Therefore, to make our barotropic model even more realistic, we need a function that smoothly joins two adjacent zones having otherwise a jump discontinuity in density. This must be done in a self-consistent manner and by simulating different levels of mixing between the two zones.

In the fiducial barotropic model of Section \ref{sec:fidmod}, we have a jump discontinuity in the density at the boundary between convective and radiative core. We can smoothly join these two zones using a sigmoid function $S(P)$ in the $\ln P - \ln \rho$ plane. This means that $\rho \equiv \rho(\rho_I, \rho_R, S)$, where $\rho_I(r)$ is the density profile in the convective core and $\rho_R(r)$ the density profile in the radiative part of the star.
A convenient choice for $S(P)$ is that it vanishes in the radiative part and becomes one in the convective core. Consequently, one can define $\rho(\rho_I, \rho_R, S) \equiv \rho(P)$ as
\begin{equation}
\label{eq:lnrho_smooth}
\ln \rho(P) := \ln \rho_R + (\ln \rho_I - \ln \rho_R) S(P),
\end{equation}
the derivative of which is
\begin{equation}
\frac{d \ln \rho}{d \ln P}  = \frac{1 - S(P)}{\gamma_R[ \rho_R( P)]} + \frac{S(P)}{\Gamma_{1,c}} + (\ln \rho_I - \ln \rho_R) \frac{d S(P)}{d \ln P}.
\end{equation}
Finally we obtain that
\begin{equation}
\label{eq:gamma_smooth}
\begin{aligned}
\frac{1}{\gamma(\rho)} & = \frac{1 - S[ P( \rho)]}{\gamma_R \{  \rho_R[ P( \rho)] \} } + \frac{S[ P( \rho)]}{\Gamma_{1,c}} + \\ & 
+ \{ \ln \rho_I[ P( \rho)] - \ln \rho_R[ P( \rho)] \} \frac{d S[ P( \rho)]}{d \ln P},
\end{aligned}
\end{equation}
with $P(\rho)$ obtained by inverting equation \ref{eq:lnrho_smooth} numerically. Therefore, the new stellar structure is obtained by solving equation \ref{eq:mass_density_diffeq} with the new $P(\rho)$ and $\gamma(\rho)$ functions obtained from equations \ref{eq:lnrho_smooth} and \ref{eq:gamma_smooth}, respectively.
As expected, $\lim_{S \to 0} \gamma = \gamma_R$, $\lim_{S \to 1} \gamma = \Gamma_{1,c}$, and $\lim_{S \to 1/2} \gamma$ has a local minimum whose value depends on the functional form of $S(P)$.

From now on we decide to use the error function as a functional form of $S(P)$:
\begin{equation}
\begin{dcases}
\label{eq:sigmoid_def}
S(P) := \frac{1 + \mathrm{erf}[\alpha(\ln P - \ln P_0)] }{2} \\
\frac{d S(P)}{d \ln P} = \frac{\alpha}{\sqrt{\pi}} \exp{[-\alpha^2(\ln P - \ln P_0)^2]}, 
\end{dcases}
\end{equation}
where $\alpha \ge 0$ is a constant whose increasing value makes the slope of $S(P)$ steeper, and $P_0$ is the pressure at the boundary between convective and radiative core in the fiducial model of Section \ref{sec:fidmod}.
With such choises, the higher is $\alpha$, the lower is the amount of mixing between the two consecutive zones. Indeed, $S(P)$ in equation \ref{eq:sigmoid_def} tends to a Heaviside function for $\alpha \to \infty$, thus, we tend to the fiducial model of Section \ref{sec:fidmod}, while for $\alpha = 0$ we have a fixed $S(P) = 1/2$ that corresponds to the maximum mixing. 
Furthermore, this choice for $P_0$ results in $\rho_I(P_0) \equiv \rho^-$, $\rho_R(P_0) \equiv \Lambda \rho^-$ and 
\begin{equation}
\lim_{P \to P_0} \frac{1}{\gamma} =  \frac{1}{2\gamma_R(\Lambda \rho^-) } + \frac{1}{2\Gamma_{1,c}} + \frac{\alpha}{\sqrt{\pi}} \ln \left( \frac{1}{\Lambda} \right).
\end{equation}
In summary, in the limit for $\alpha \to \infty$, equation \ref{eq:gamma_smooth} tends to
\begin{equation}
\label{eq:gamma_smooth_high_alpha}
\lim_{\alpha \to \infty} \gamma(\rho) = 
\begin{dcases}
\Gamma_{1,c} , \quad P > P_0 \\
0 , \quad P = P_0 \\
\gamma_R(\rho) , \quad P < P_0 ,
\end{dcases}
\end{equation}
and equation \ref{eq:lnrho_smooth} tends to
\begin{equation}
\label{eq:gamma_smooth_high_alpha}
\lim_{\alpha \to \infty} \rho(P) = 
\begin{dcases}
\rho_I(P) , \quad P > P_0 \\
\rho_R(P) , \quad P < P_0,
\end{dcases}
\end{equation}
which means that we tend to the fiducial model of Section \ref{sec:fidmod} as expected.

To obtain the inverse of $\rho(P)$ we need to evaluate $\ln \rho_I (P)$ and $\ln \rho_R (P)$. From Section \ref{sec:firstzone} we obtain
\begin{equation}
\ln \rho_I (P) = \ln \rho_c + \frac{\ln P - \ln P_c}{\Gamma_{1,c}},
\end{equation}
but $\ln \rho_R (P)$ does not have an analytical solution. However, 
\begin{equation}
\label{eq:approx_second_zone}
\ln P_R(\rho) \approx \ln P_0 + \frac{a}{K}\frac{\rho^K - \rho^K_+}{\rho^K_+}
\end{equation}
is a very good approximation of equation \ref{eq:ln_beta_second_zone} when $\rho \gtrsim 0.24 \Lambda \rho^-$, because in that range their relative difference is $|1 - P_\mathrm{true} / P_\mathrm{approx} | < 10^{-10} \%$. Therefore, the inverse is well approximated by
\begin{equation}
\label{eq:approx_second_zone_2}
\ln \rho_R(P) \approx \ln (\Lambda \rho^- ) + \frac{\ln \left[ 1 + \frac{K}{a} \ln \left( \frac{P}{P_0} \right) \right] }{K}
\end{equation}
when $\rho_R \gtrsim 0.24 \Lambda \rho^-$. Finally we obtain that
\begin{equation}
\label{eq:smooth_gamma_rho_1}
\begin{aligned}
\frac{1}{\gamma(\rho)} & \approx \frac{1 - S[P(\rho)]}{ a + K [\ln P(\rho) - \ln P_0] } + \frac{S[P(\rho)]}{\Gamma_{1,c}} + \\ &
+ \ln \left \{ \frac{\rho_c}{\Lambda \rho^-}  \left[ \frac{P(\rho)}{P_c} \right]^{\frac{1}{\Gamma_{1,c}}} \left[ \frac{a}{a + K(\ln P - \ln P_0) } \right] ^\frac{1}{K}  \right \} \frac{d S[P(\rho)]}{d \ln P}
\end{aligned}
\end{equation}
and
\begin{equation}
\label{eq:smooth_gamma_rho_2}
\begin{aligned}
\ln \rho(P)  & \approx \ln (\Lambda \rho^- ) + \frac{\ln \left[ 1 + \frac{K}{a} \ln \left( \frac{P}{P_0} \right) \right] }{K} + \\ &
+\ln \left \{ \frac{\rho_c}{\Lambda \rho^-}  \left( \frac{P}{P_c} \right)^{\frac{1}{\Gamma_{1,c}}} \left[ \frac{a}{a + K(\ln P - \ln P_0) } \right] ^\frac{1}{K}  \right \} S(P) \\
\end{aligned}
\end{equation}
when $\rho_R \gtrsim 0.24 \Lambda \rho^-$, and from a numerical evaluation of $P (\rho)$ we obtain $\gamma(\rho)$. 

For $|\alpha (\ln P - \ln P_0)| >> 1$ we have $\gamma \approx \Gamma_{1,c}$ for $\rho \approx \rho^-$, and $\gamma \approx \gamma_R$ for $\rho \approx \rho^+$.
Note that to a good approximation
\begin{equation}
\label{eq:approx_024_and_alpha_1}
\begin{aligned}
& P(\rho = 0.24 \Lambda \rho^-) \approx P_R(\rho_R = 0.24 \Lambda \rho^-)  \quad \mathrm{for} \quad \alpha \ge 1.885 \\
& \gamma(\rho = 0.24 \Lambda \rho^-) \approx \gamma_R(\rho_R = 0.24 \Lambda \rho^-)  \quad \mathrm{for}  \quad \alpha \ge 2.074 ,
\end{aligned}
\end{equation}
since the ratio of the two sides of the equation deviates from unity by less than $10^{-6} \%$. This means that the approximation in equation \ref{eq:approx_second_zone_2}, which is valid when $\rho_R \gtrsim 0.24 \Lambda \rho^-$, is also valid when $\rho \gtrsim 0.24 \Lambda \rho^-$ if $\alpha \ge 2.074$. Moreover, a consequence of equation \ref{eq:approx_024_and_alpha_1} is that $\gamma (\rho \leq 0.24 \Lambda \rho^-) \approx \gamma_R (\rho_R \leq 0.24 \Lambda \rho^-)$ and $P (\rho \leq 0.24 \Lambda \rho^-) \approx P_R (\rho_R \leq 0.24 \Lambda \rho^-)$ if $\alpha \ge 2.074$.
Therefore, we decide to solve equations \ref{eq:smooth_gamma_rho_1} and \ref{eq:smooth_gamma_rho_2} when $\rho \ge 0.24 \Lambda \rho^-$, and when $\rho < 0.24 \Lambda \rho^-$ (i.e. until the end of the second zone) we solve $\gamma \approx \gamma_R$ using equation \ref{eq:gamma_second_zone}
and
\begin{equation}
\begin{aligned}
 & \ln P(\theta) \approx \ln P_R = \ln P_1 + \int_{\ln \theta_1}^{\ln \theta}\gamma_R(t) \, dt \\
 &= \ln P_1 +
\begin{dcases}
 A \sqrt{\frac{\pi}{2}} \sigma\left[ \mathrm{erf} \left( \frac{\ln \theta - \ln \theta_P}{ \sqrt{2} \sigma} \right)  + \mathrm{erf} \left( \frac{\ln \theta_P - \ln \theta_1}{ \sqrt{2} \sigma} \right)  \right] + \\
 + \frac{a}{K} \left( \theta^K - \theta_1^K  \right)  , \quad \theta_P \le \theta < \theta_1 \\\\
A \sqrt{\frac{\pi}{2}} \sigma\left[ \mathrm{erf} \left( \frac{\ln \theta_P - \ln \theta_1}{ \sqrt{2} \sigma} \right)  \right]+ \frac{a}{K} \left( \theta_P^K - \theta_1^K   \right) + \\
+ [A + a \theta_P^K - (a') \theta_P^{K'}] \sqrt{\frac{\pi}{2}} (\sigma')  \mathrm{erf} \left( \frac{\ln \theta - \ln \theta_P}{ \sqrt{2} \sigma'} \right) + \\
+ \frac{a'}{K'} \left( \theta^{K'} -  \theta_P^{K'}  \right), \quad  \theta < \theta_P \, \, .
\end{dcases}
\end{aligned}
\end{equation}
where $\theta := \rho / (\Lambda \rho^-)$, $\theta_1 := 0.24$ and $\ln P_1 := \ln P(\theta_1)$, which is found using equations \ref{eq:smooth_gamma_rho_1} and \ref{eq:smooth_gamma_rho_2}. Finally, we choose $\Gamma_1$ such that 
\begin{equation}
\Gamma_1 =
\begin{dcases}
\Gamma_{1,c} \quad \mathrm{for} \quad \rho \ge \rho^- \\
\left( \frac{5}{3} - \Gamma_{1,c} \right) \frac{\ln \left(\rho\right) - \ln (\Lambda  \rho^-)}{\ln (\Lambda)} + \frac{5}{3} \quad \mathrm{for} \quad \Lambda \rho^- < \rho < \rho^- \\
\frac{5}{3} \quad \mathrm{for} \quad \rho \le \Lambda \rho^-   .
\end{dcases}
\end{equation}

\begin{figure}[htbp]
\centering
\includegraphics[width=\columnwidth]{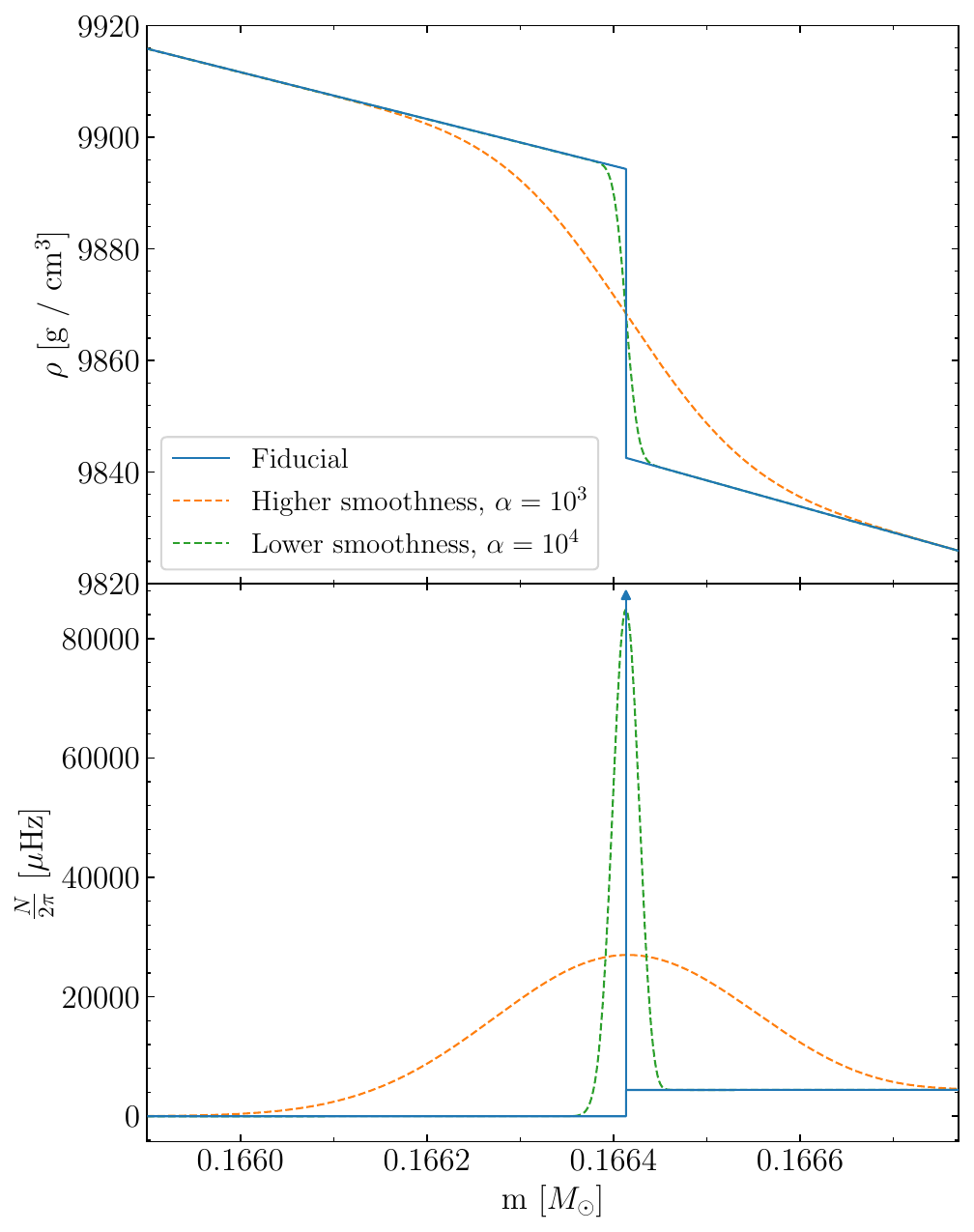} 
\caption{Comparison between three different density profiles (top panel) at the boundary between the convective and the radiative core and their corresponding Brunt-Väisälä frequencies (bottom panel) as functions of internal mass. In particular, we compare the fiducial barotropic model (blue) with two models that smoothly join the two zones (orange and green). The orange line incorporates a lower $\alpha$ value than the green line. The blue arrow corresponds to a $\delta$-distribution. The FWHM of the peaks in $N$ presented here are significantly lower than $0.1H_p$.}
\label{fig:smooth_core_boundary} 
\end{figure}
With this formalism we can explore different behaviours at the boundary between convective and radiative core by varying the parameters $\Lambda$ and $\alpha$. In the top panel of Figure \ref{fig:smooth_core_boundary}, we illustrate three scenarios: one with a jump discontinuity in density (blue lines), and two models that share the same $\Lambda$ as the blue case but differ in $\alpha$ values (orange and green lines). The bottom panel displays the corresponding Brunt-Väisälä frequencies as functions of internal mass, highlighting that a higher $\alpha$ leads to a sharper bell-shaped structure. Furthermore, the same formalism can be applied to smoothly connect two adjacent zones that would otherwise exhibit a jump discontinuity in density within the radiative core. In this context, the primary distinction lies in the fact that $\gamma_I(\rho) \neq \Gamma_{1,c}$, and the validity of the approximation presented in the equations must be verified.

\section{Results}
\label{sec:results}
In this section we explain the main results concerning our fiducial barotropic model (Section \ref{sec:fiducial_model}) in comparison with a different boundary of the convective core (Section \ref{sec:glitch_boundary_cc}), different glitches in the radiative core (Section \ref{sec:glitch_translated_discont}), and different degree of jump discontinuities near the boundary of the convective core (Section \ref{sec:diff_jump_discont_core}). It is important to note that some of these glitches are primarily used as test cases, as within the observable region of the oscillation spectrum they would appear too smooth to be identified as glitch signatures.

We compute the adiabatic eigenfrequencies, normalised inertia \citep[$E_{\rm norm}$, see the definition in e.g.][]{Aerts2010} and eigenfunctions of radial ($\ell=0$) and non-radial ($\ell=1-3$) modes using the code \texttt{GYRE} \citep[version 6.0.1,][]{Townsend2013,Townsend2018,Goldstein2020}, but we verify independently the calculations with the tool \texttt{LOSC} \citep{Scuflaire2008b} in the case of the fiducial model. We also compute the uncoupled g-modes, called $\gamma$-modes, through the prescriptions provided by \citet{Ong2020} and included in \texttt{GYRE}. As will be clear in the subsequent sections, this simplifies the detection of buoyancy glitches.
Finally, we simulate 4-year-long \kepler observations (lightcurves and power spectral densities) with the code \texttt{AADG3} \citep[AsteroFLAG Artificial Dataset Generator, version 3.0.2;][and references therein]{Ball2018}, with information on mode lifetimes taken from the work of \citet[][]{Vrard2018} on CHeB stars in the \kepler field.

\subsection{Fiducial model}
\label{sec:fiducial_model}
In this section, we highlight the results of the fiducial barotropic model of Section \ref{sec:fidmod}.
\begin{figure*}[htbp]
\centering
\includegraphics[width=\columnwidth]{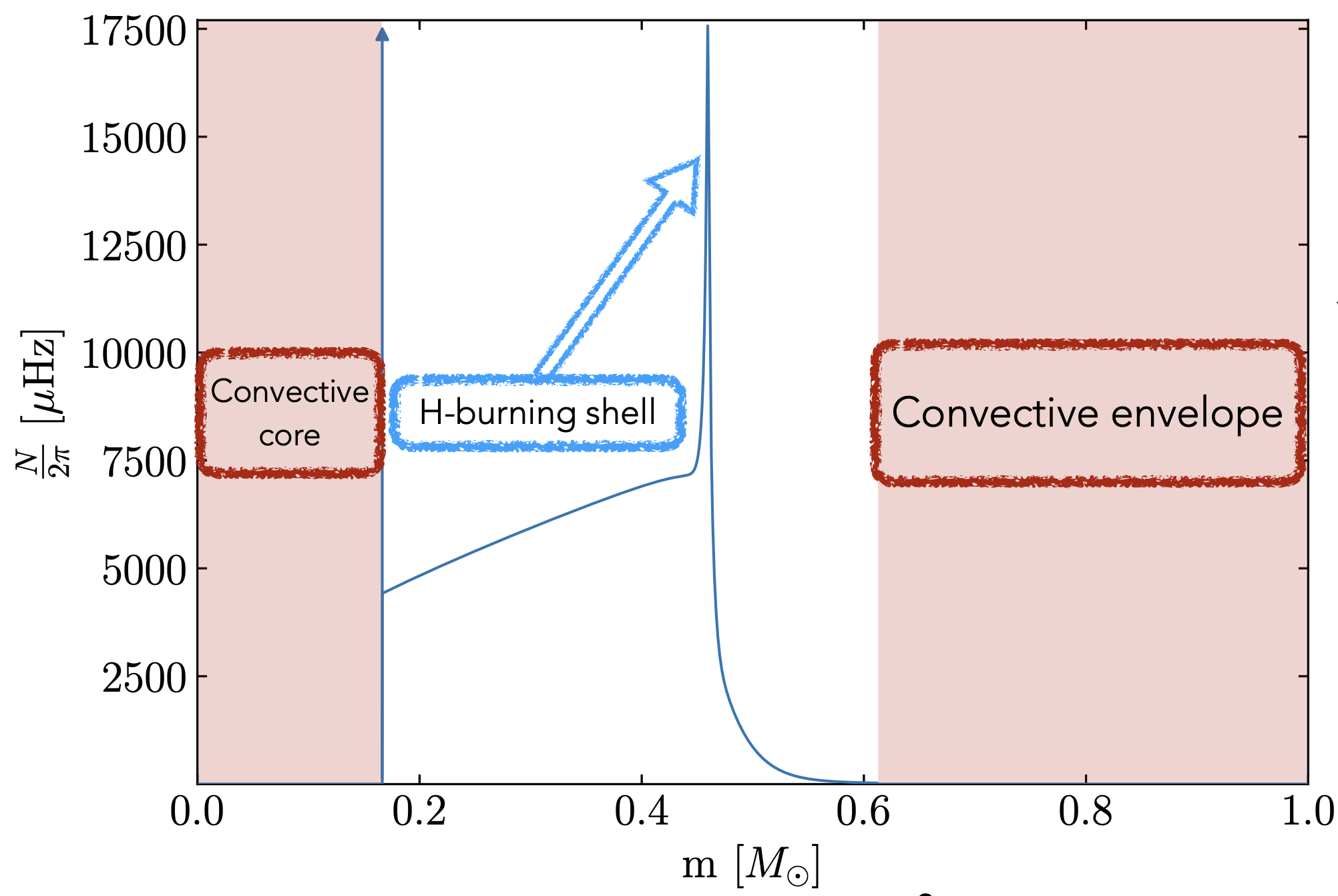} 
\includegraphics[width=\columnwidth]{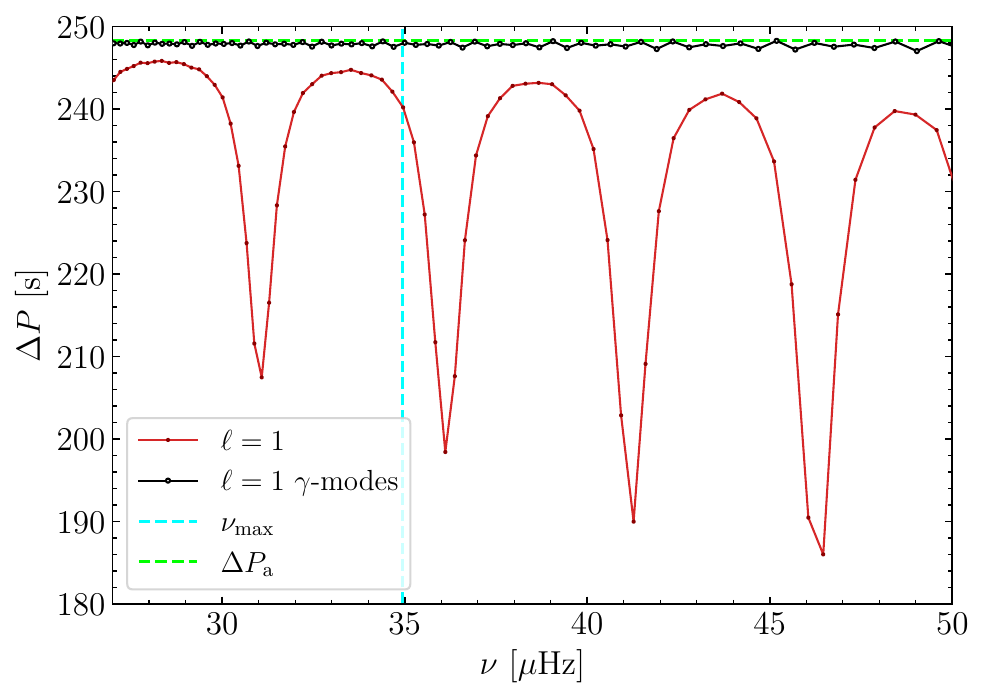} 
\includegraphics[width=0.85\linewidth]{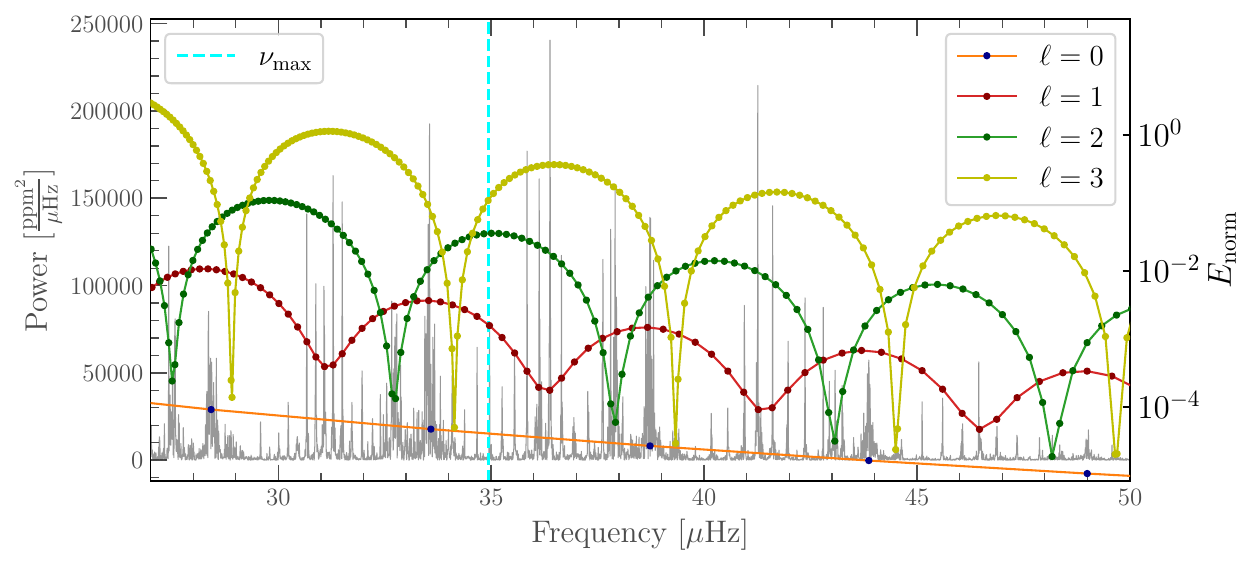} 
\caption{This figure displays the Brunt-Väisälä frequency as a function of internal mass (top left), period spacing as a function of eigenfrequencies in the observable region of the oscillation spectrum (top right), and the corresponding simulated PSD (bottom) for the fiducial model discussed in Section \ref{sec:fiducial_model}. The cyan dashed line indicates the \numax of the reference \texttt{CLES} model. Along with the simulated PSD, we also show the normalised inertia of radial and non-radial modes (coloured dots connected by coloured lines).}
\label{fig:fiducial_model_brunt_inertia} 
\end{figure*}
In Figure \ref{fig:fiducial_model_brunt_inertia} we show the Brunt-Väisälä frequency as a function of the internal mass, with a specification for the position of the convective core, the hydrogen-burning shell and the convective envelope (top panel). In the boundary between the convective and radiative core there is a change in $\Gamma_1$, a change in $\gamma(\rho)$ and a jump discontinuity in density. This implies that $N$ is a $\delta$-distribution (blue arrow in the figure) at the position corresponding to the boundary. In the BCE, the density profile is non-differentiable due to a "jump" in $\gamma(\rho)$, thus here $N$ has a jump discontinuity.

As discussed at the beginning of Section \ref{sec:results}, we assess our model by computing adiabatic eigenfrequencies that serve to simulate 4-year-long \kepler observations. The corresponding power spectral density (PSD) as a function of frequency is shown in the bottom panel of Figure \ref{fig:fiducial_model_brunt_inertia}. This panel includes a cyan dashed line representing the \numax, derived from the reference \texttt{CLES} model, alongside the normalised inertiae as a function of eigenfrequencies for modes with $\ell =0, 1, 2, \text{ and } 3$. As expected \citep[e.g.][]{Montalban2010,Mosser2011a,Mosser2012b}, the radial modes, along with the minima of inertia for the non-radial modes, are evenly spaced in frequency with a $\langle \Delta \nu \rangle \approx 5 \mu$Hz. Additionally, a "forest" of dipole mixed modes appears between two consecutive radial modes, a characteristic observed in actual low-mass CHeB stars. We further expect that, in these stars, the period spacing ($\Delta P$) of the observable dipole mixed modes approaches the asymptotic value ($\Delta P_\mathrm{a}$), which is notably higher than that of RGB stars \citep[e.g.][]{Montalban2010,Bedding2011}. Our fiducial model confirms this expectation,
as illustrated in the upper right panel of Figure \ref{fig:fiducial_model_brunt_inertia}. It is apparent that as the frequency decreases, $\Delta P$ approaches the asymptotic value of $\Delta P_\mathrm{a} = 248.34 \, \mathrm{s}$ due to the increasing radial order of the g-modes. Moreover, each minimum in $\Delta P$ corresponds to a minimum of inertia, linked to a pressure-dominated mode, and the higher the frequency the lower the $\Delta P$. We also calculate the coupling parameter ($q$) at \numax using the prescriptions of \citet{Takata2016a}, yielding $q \approx 0.24$, which aligns with the observational findings \citep[e.g.][]{Vrard2016,Mosser2017}.

In the hydrogen-burning shell, we observe that the Brunt-Väisälä frequency is non-differentiable at the point where $\theta = \theta_P$, a condition resulting from the specific choice of $\gamma(\rho)$ employed in equation \ref{eq:gamma_second_zone}. This non-differentiability has the potential to introduce an artificial glitch in the period spacing of the mixed modes, as noted in previous studies \citep[e.g.][]{Cunha2015}. Therefore, it warrants careful examination.
In Appendix \ref{app:artif_glitches_fiducial_model}, we argue that the small deviations in $\Delta P_\mathrm{a}$ noted in our model are likely attributable to numerical inaccuracies in the computation of eigenfrequencies rather than the non-differentiable point itself. 
Notably, the maximum observed peak-to-peak relative difference in period spacing of the $\gamma$-modes is at most 0.2 \%, which is significantly lower than the typical observational uncertainties \citep[e.g., $\approx 3$ s in the asymptotic period spacing of CHeB stars in the \kepler database,][]{Vrard2016,Vrard2022}. This indicates that the glitches identified in our model do not compromise the interpretation of observed power spectral densities. Additionally, this suggests that the interpretation of other glitches, which will be discussed in subsequent sections, are unlikely to be affected by this specific instance of non-differentiability or by numerical errors. Furthermore, it is important to point out that the hydrogen-burning shell, which is characterised by a half width at half maximum (HWHM) of approximately 0.3 $H_P$ in our model, is incapable of producing a glitch signature in the eigenfrequencies. Any potential evidence of such a signature may only be detected in the very early stages of the CHeB phase, particularly when the hydrogen-burning shell is relatively thin.

\subsection{Different boundary of the convective core}
\label{sec:glitch_boundary_cc}
\begin{figure}[htbp]
\centering
\includegraphics[width=\columnwidth]{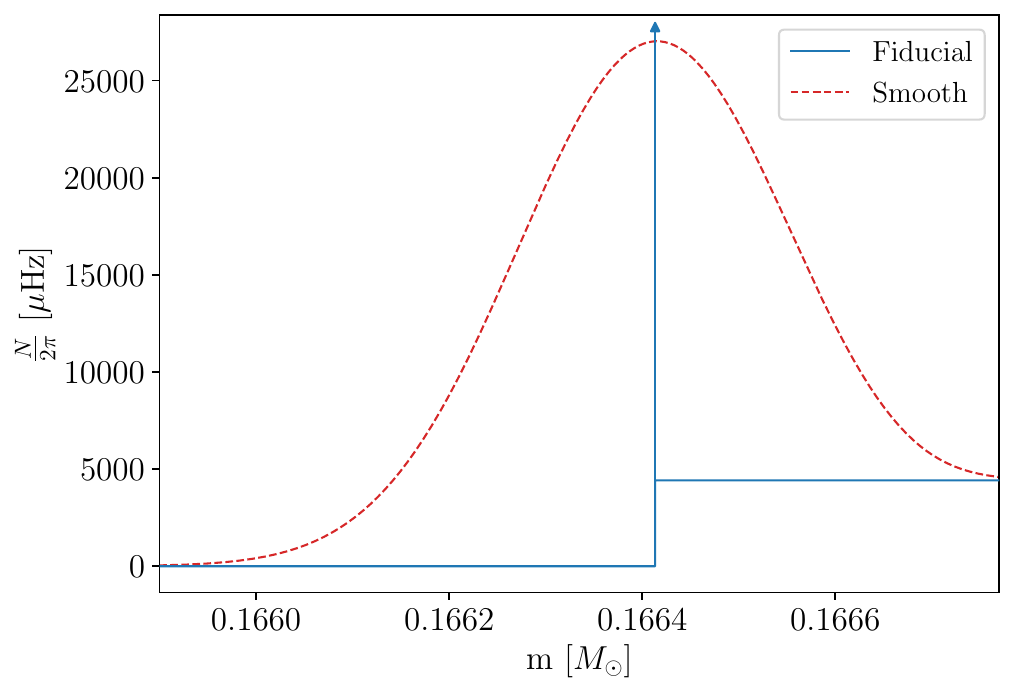} 
\includegraphics[width=\columnwidth]{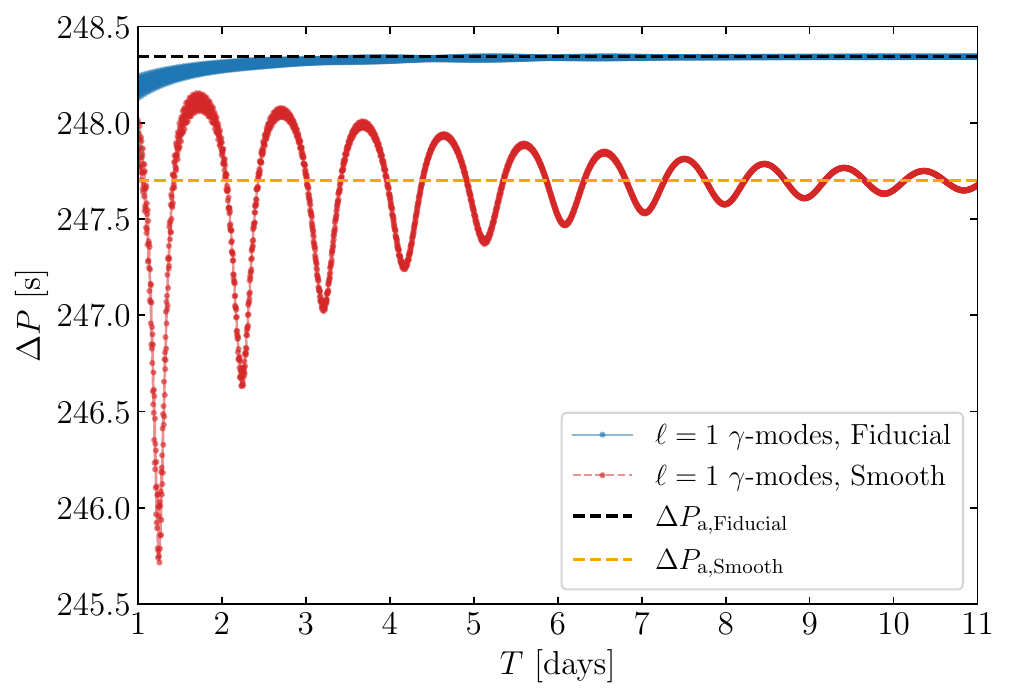} 
\caption{Comparison of $N$ profiles as functions of internal mass (top panel) and period spacings of the $\gamma$-modes as functions of the eigenperiods (bottom panel). The blue line represents the fiducial model, while the red model features a bell-shaped structure in the $N$ profile at the boundary between the convective and radiative core (see Section \ref{sec:smoothdiscont}). The periodicity of the glitch corresponds to the location of the bell-shaped structure. The half width at half maximum of the bell-shaped structure presented here is $0.0012 H_P$.}
\label{fig:smooth_core_boundary_inertia} 
\end{figure}
In this section, we explore the impact on frequencies of a model similar to the fiducial barotropic model, but with a smooth transition in the density profile between the convective and radiative core (Section \ref{sec:smoothdiscont}). In Figure \ref{fig:smooth_core_boundary_inertia} we show in the top panel its $N$ profile (red line) as a function of the internal mass in comparison with the fiducial model (blue line). This new profile is very similar to the fiducial model, but it contains a bell-shaped structure centred at the boundary instead of a $\delta$-distribution. We expect a value of the asymptotic period spacing of the new model $\Delta P_\mathrm{a, Smooth}$ lower than the one of the fiducial model $\Delta P_\mathrm{a, Fiducial}$, because the new model has a slighlty greater g-cavity than the fiducial model and it has higher values of $N$ near the boundary. Moreover, we expect that such bell-shaped structure in the $N$ profile becomes a glitch, which creates a sinusoidal behaviour of $\Delta P$ around $\Delta P_\mathrm{a, Smooth}$ with a decreasing amplitude at increasing period, and dips in the period spacing that are evenly spaced in period, with a periodicity compatible with the buoyancy radius of the glitch \citep{Cunha2019,Cunha2024}. To verify that the model behaves as predicted by theory, we calculate eigenmodes also outside the observable region of the spectrum.

In the bottom panel we show the period spacings of the $\gamma$-modes as functions of the eigenperiods for this new model (in red) and the fiducial model (in blue). As expected, there is a very small difference between the two asymptotic period spacings, that is, $\Delta P_\mathrm{a, Smooth} \approx \Delta P_\mathrm{a, Fiducial} - 0.65$ s. Moreover, we observe in the new model a periodicity of approximately 23 hours with a decreasing amplitude at increasing eigenperiods. When expressed in terms of radial orders, this same periodicity corresponds to $\Delta n \approx 334$, which aligns with the expected value derived from applying equation \ref{eq:Tau_signal_Dk} to the new model. According to the same equation, we also infer that this glitch signature correlates to a structural variation situated at $\Phi (r_\mathrm{glitch}) \approx 0.0030$. This finding is consistent with the normalised buoyancy radius of the bell-shaped structure in $N$, which peaks at $\Phi \approx 0.0015$ and possesses a FWHM of 0.0023. To further support the relation between the bell-shaped structure in $N$ and the observed glitch signature, we compare the Brunt-Väisälä length scale ($H_N$) with the local wavelength of g-modes ($\lambda_g$, as discussed in Section \ref{sec:intro}) near \numax.
\begin{figure}[htbp]
\centering
\includegraphics[width=\columnwidth]{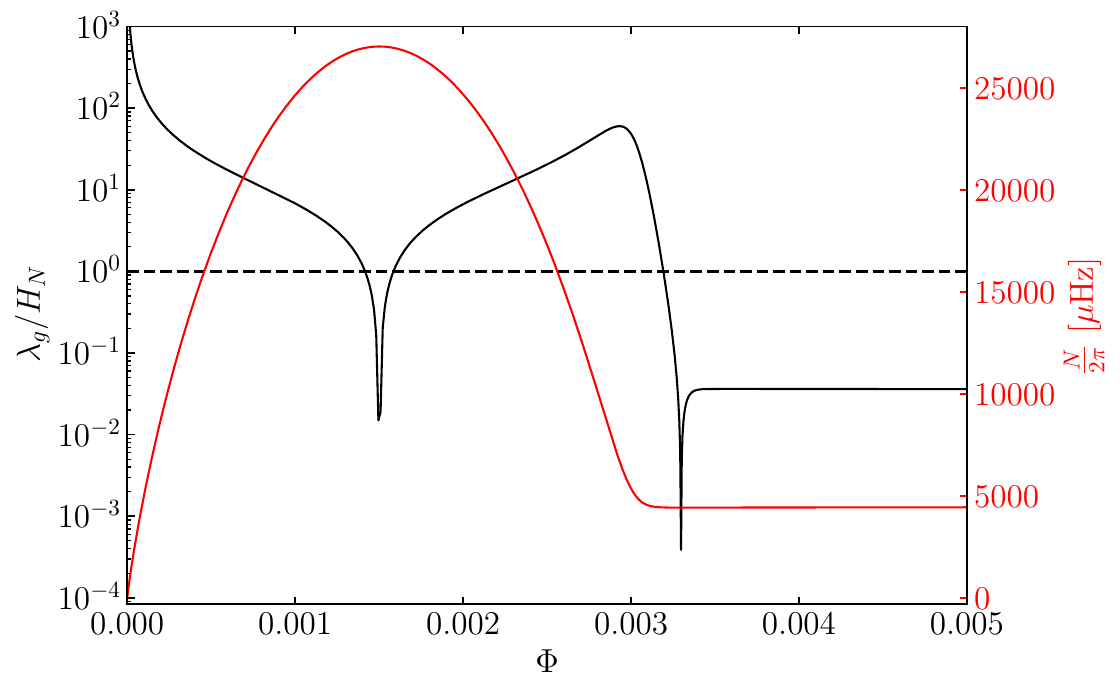} 
\caption{In black, we show the ratio of the local wavelength of g-modes ($\lambda_g$) near \numax to the Brunt-Väisälä length scale ($H_N$) as a function of the normalised buoyancy radius ($\Phi$) for a model characterised by a bell-shaped structure in the $N$ profile at the boundary between the convective and radiative core (refer to Section \ref{sec:smoothdiscont}). The dashed black line indicates the threshold where $\lambda_g/H_N = 1$. Additionally, we present the $N$ profile (red line) as a function of $\Phi$.}
\label{fig:HN_lambda_g_smooth_centre} 
\end{figure}
We find that $H_N$ of the bell-shaped structure is significantly smaller than $\lambda_g$ at $\Phi (r_\mathrm{glitch}) \approx 0.0030$ (see Figure \ref{fig:HN_lambda_g_smooth_centre}).
It is important to highlight that within the observable region of the oscillation spectrum, this periodic component would appear too smooth to be identified as a glitch signature. Nonetheless, it serves as a useful case study for testing, and we cannot rule out the possibility that signatures of convective boundary mixing may be observed in more evolved CHeB stars, where glitches at the boundary between the convective and radiative core could potentially be more pronounced (see Section \ref{sec:diff_jump_discont_core}). Finally, the small glitch explained in Section \ref{sec:fiducial_model} adds to the above glitch, generating relative fluctuations much lower than 1 \% around the main behaviour of the period spacing.

\subsection{Glitches in the radiative core}
\label{sec:glitch_translated_discont}
In this section, we explore the impact on frequencies of two different types of glitches in the radiative core. In Section \ref{sec:translated_discont} we study the impact of a jump discontinuity in density, whereas in Section \ref{sec:Smooth_translated_discont} we smooth the same discontinuity using the method of Section \ref{sec:smoothdiscont}.

\subsubsection{$\delta$-distribution structural glitch}
\label{sec:translated_discont}
\begin{figure}[htbp]
\centering
\begin{subfigure}[b]{\linewidth}
\centering
\includegraphics[width=\linewidth]{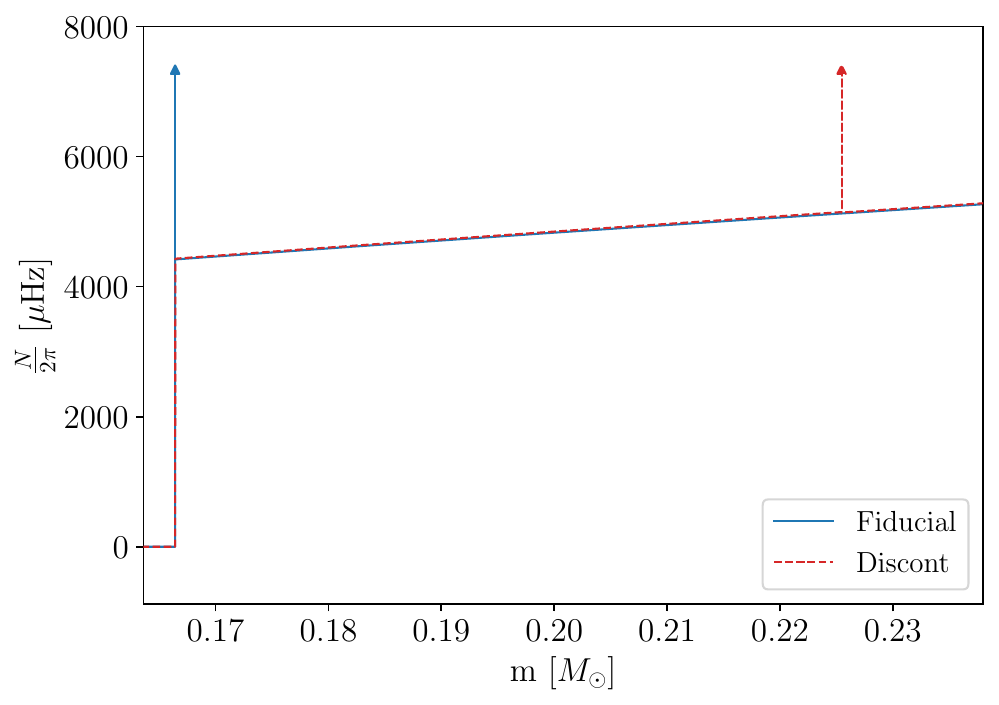} 
\end{subfigure}
\begin{subfigure}[b]{\linewidth}
\centering
\includegraphics[width=\linewidth]{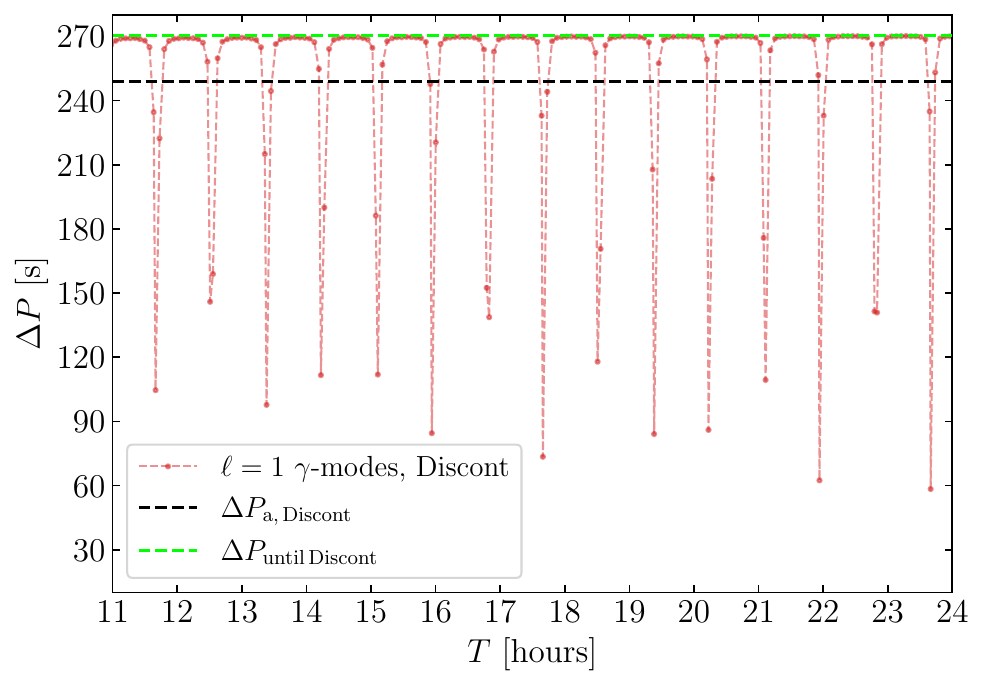} 
\end{subfigure}
\caption{This figure presents a comparison of the $N$ profiles as functions of internal mass (top panel) and the period spacings of the $\gamma$-modes as functions of the eigenperiods (bottom panel). The blue line represents the fiducial model, while the red model incorporates a $\delta$-distribution glitch in the $N$ profile located within the radiative core. The fiducial model is not displayed in the bottom panel, as it would be indistinguishable from the black line, which represents the $\Delta P_\mathrm{a}$ of the red model. Notably, the periodicity in the bottom panel corresponds to the location of the $\delta$-distribution, and the maximum period spacings measured closely match $\Delta P_\mathrm{a, until \, Discont}$. This value reflects the asymptotic $\Delta P$ we would obtain if the g-cavity extends from the position of the glitch to the outer boundary.}
\label{fig:translated_discont} 
\end{figure}
In Figure \ref{fig:translated_discont}, the top panel displays the $N$ profile as a function of internal mass for a model similar to that presented in Appendix \ref{sec:no_discont}. This model features a continuous, albeit non-differentiable, density function at the boundary between the convective and radiative core (see Appendix \ref{sec:no_discont} for more details), and it introduces a new jump discontinuity in density that is compatible with a subflash event occurring within the radiative core or with the chemical gradient found at the end of an overshoot region that possesses radiative thermal stratification. This conclusion is drawn from the observation that the jump present in the reference \texttt{CLES} model at the outer edge of the helium core closely resembles that at the boundary between the convective and radiative core. We have selected a position of $r = 0.03 \, \mathrm{R}_\odot$ for the jump, as it correlates well with a main flash event identified in other evolutionary codes \citep{Kippenhahn2012}. The total stellar radius of our barotropic models is $9.9457 \, \mathrm{R}_\odot$, which provides a reference scale for interpreting the position discussed above. The resulting $N$ profile (red line) closely resembles the fiducial model (blue line); however, the red model is characterised by a jump discontinuity at the boundary between the convective and the radiative core rather than a $\delta$-distribution. Moreover, a $\delta$-distribution is observed at the location of the jump discontinuity in density within the radiative core, with the bell-shaped peak associated with the hydrogen-burning shell occurring at a lower radius compared to the fiducial model. Overall, the integral $\int_{r_1}^{x} N / r \, dr$ at a fixed position $x \leq r_2$ in this new model resembles that of the Appendix \ref{sec:no_discont} mentioned earlier. However, the jump discontinuity in density within the radiative core leads to a reduction in $\int_{r_1}^{x} N / r \, dr$ for positions above the discontinuity when compared to the other case. This cumulative effect drives the value of the integral $\int_{r_1}^{r_2} N / r \, dr$ below that of the fiducial model. Consequently, we expect a value of the asymptotic period spacing $\Delta P_\mathrm{a, Discont}$ higher than $\Delta P_\mathrm{a, Fiducial}$.
Furthermore, we expect that the $\delta$-distribution in the $N$ profile will manifest as a glitch, resulting in periodic deviations of $\Delta P$ relative to $\Delta P_\mathrm{a, Discont}$ \citep[e.g.][]{Cunha2015}. These deviations will exhibit a decreasing amplitude with increasing frequency\footnote{We would like to remind the reader that the opposite trend is expected when the structural glitch has a non-zero width value, as explained by \citet{Cunha2015,Cunha2019}.}, creating evenly spaced dips in period that align with the periodicity associated with the normalised buoyancy radius of the glitch. Additionally, some modes are expected to become trapped in regions close to the density discontinuity as a result of the abrupt change in $N$. To verify that the model behaves as predicted by theory, we calculate eigenmodes also outside the observable region of the spectrum.

In the bottom panel, the period spacing of the $\gamma$-modes as a function of the eigenperiods for this new model. The fiducial model is not displayed in the bottom panel, as it would be indistinguishable from $\Delta P_\mathrm{a, Discont}$, which represents the $\Delta P_\mathrm{a}$ of the red model. As expected, there is a small difference between the two asymptotic period spacings, specifically $\Delta P_\mathrm{a, Discont} \approx \Delta P_\mathrm{a, Fiducial} + 0.46$ s. Moreover, the period spacing exhibits a periodicity of about 51 minutes; when expressed in terms of radial orders, this periodicity translates to $\Delta n \approx 12.4$, in accordance with the expected value derived from the application of equation \ref{eq:Tau_signal_Dk} to the model. This analysis also suggests that the glitch signature corresponds to a structural variation situated at $\Phi (r_\mathrm{glitch}) \approx 0.080$, which is compatible with the normalised buoyancy radius of the $\delta$-distribution. Notably, the maximum period spacings measured closely match $\Delta P_\mathrm{a, until \, Discont}$. This value reflects the asymptotic $\Delta P$ we would obtain if the g-cavity extends from the position of the glitch to the outer boundary. It is evident that the inferred $\Delta P_\mathrm{a}$ from observations would resemble $\Delta P_\mathrm{a, until \, Discont}$ rather than the true $\Delta P_\mathrm{a}$, as expected \citep[e.g.][]{Cunha2015}. Finally, the small glitch explained in Section \ref{sec:fiducial_model} adds to the above glitch, generating relative fluctuations much lower than 1 \% around the main behaviour of the period spacing.

\begin{figure*}[htbp]
\centering
\includegraphics[width=0.6\linewidth]{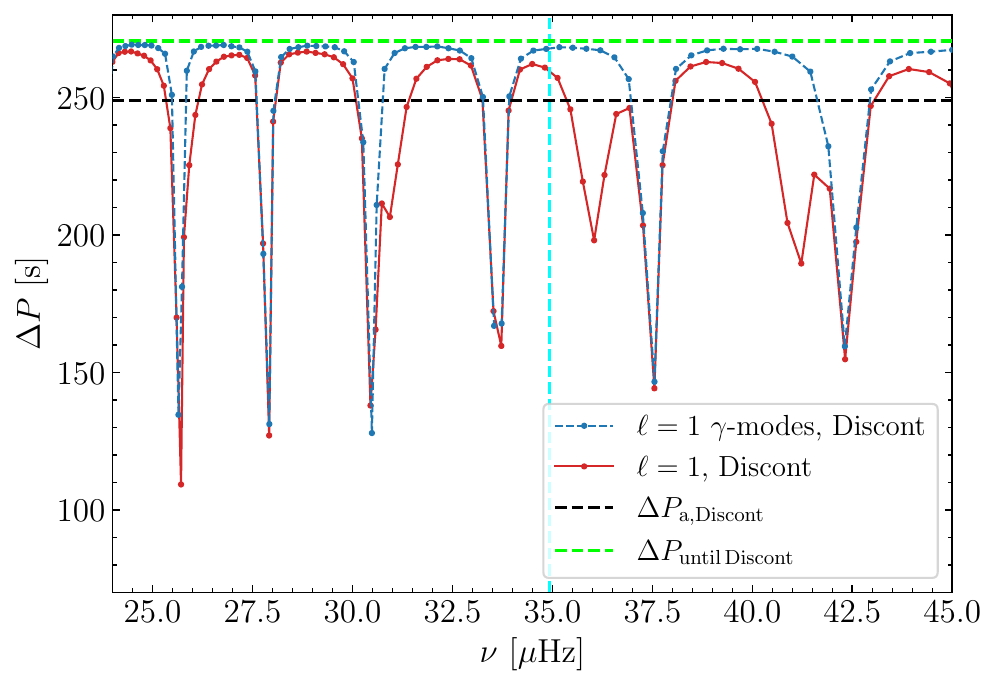} 
\includegraphics[width=0.85\linewidth]{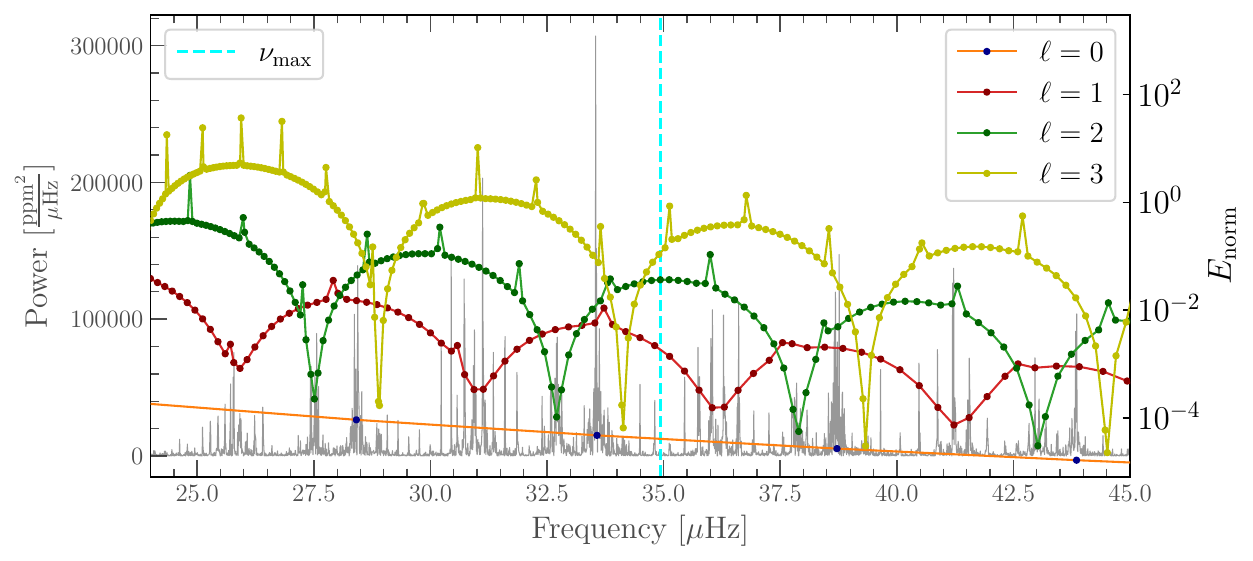}
\caption{This figure displays the period spacing (top panel) as a function of eigenfrequencies for the model discussed in Section \ref{sec:translated_discont}. We compare the $\gamma$-modes (shown in blue) with mixed dipole modes (in red) in proximity to the observable region of the oscillation spectrum, whose \numax is the cyan line. The black and lime lines are the same as Figure \ref{fig:translated_discont}. Some minima in the period spacing result from mode trapping rather than the coupling between p-modes and g-modes, and not all of these minima exhibit observable amplitudes. This is further illustrated in the bottom panel, which presents the corresponding simulated PSD along with the normalised inertia of radial and non-radial modes (coloured dots connected by coloured lines).}
\label{fig:trapping_modes_discont_translated} 
\end{figure*}
After testing our model, in the top panel of Figure \ref{fig:trapping_modes_discont_translated} we compare the period spacing of the $\gamma$-modes (in blue) with the period spacing of the mixed dipole modes (in red) in the observable region of the spectrum. Notably, some minima in the period spacing result from mode trapping rather than the coupling between p-modes and g-modes. This suggests that certain trapped modes may be detectable in actual data. However, not all of these minima show observable amplitudes, as modes with higher inertia tend to have lower amplitudes \citep[e.g.][]{Dupret2009,Grosjean2014}, leading to decreased detectability even when the period spacing is low. Consequently, the measured $\Delta P$ is likely to differ from the true value if some modes are absent. This situation becomes even clearer in the bottom panel of Figure \ref{fig:trapping_modes_discont_translated}, where we present the PSD (along with the normalised inertiae) for the model featuring a density discontinuity within the radiative core.

\subsubsection{Bell-shaped structural glitch}
\label{sec:Smooth_translated_discont}
\begin{figure}[htbp]
\centering
\begin{subfigure}[b]{\linewidth}
\centering
\includegraphics[width=\linewidth]{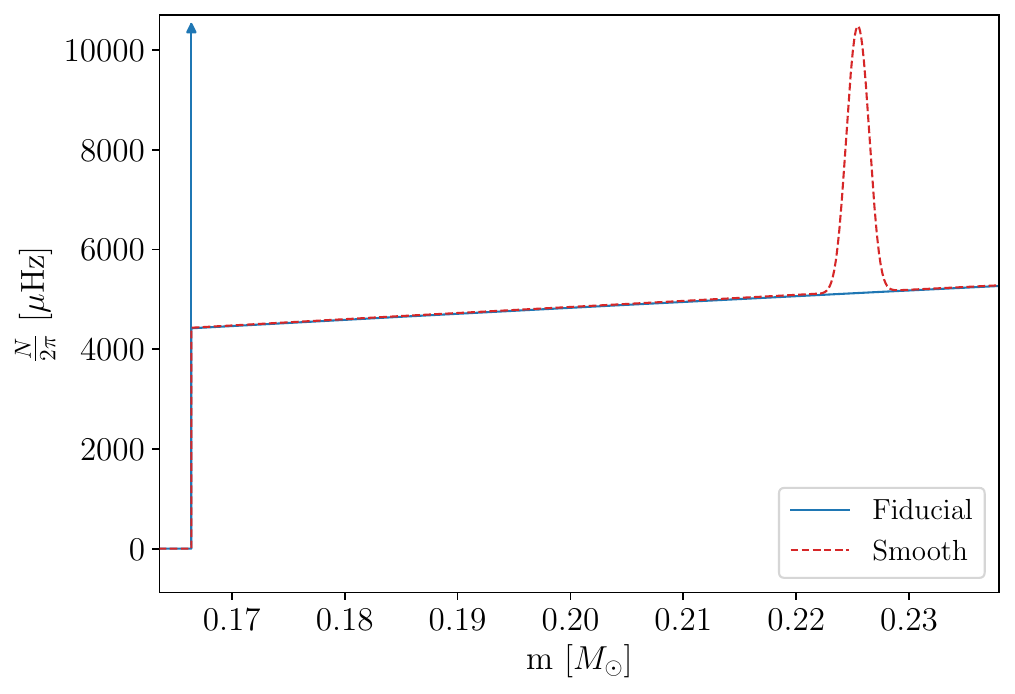} 
\end{subfigure}
\begin{subfigure}[b]{\linewidth}
\centering
\includegraphics[width=\linewidth]{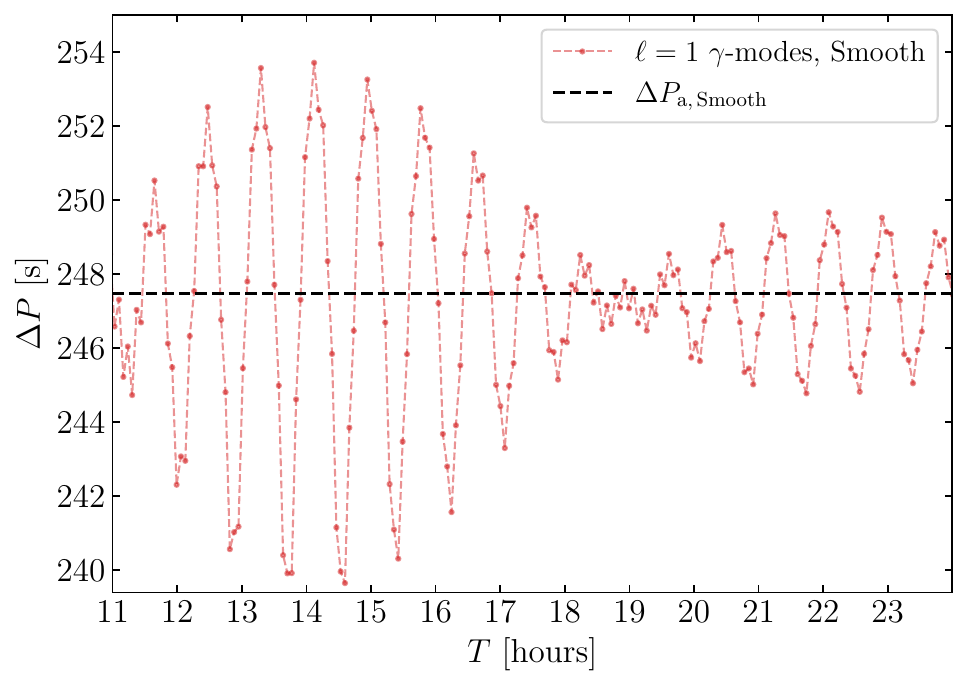} 
\end{subfigure}
\caption{Comparison of $N$ profiles as functions of internal mass (top panel) and period spacings of the $\gamma$-modes as functions of the eigenperiods (bottom panel). The blue line represents the fiducial model, while the red model features a bell-shaped structure in the $N$ profile within the radiative core. The half width at half maximum of the bell-shaped structure in $N$ presented here is $0.02H_P$.}
\label{fig:Smooth_translated_discont} 
\end{figure}
In the top panel of Figure \ref{fig:Smooth_translated_discont}, we show the $N$ profile as a function of the internal mass for a model similar to the model of Section \ref{sec:translated_discont}, but with a smooth transition in density instead of a jump discontinuity (i.e. with $\alpha = 10^2$, see Section \ref{sec:smoothdiscont}). This new model has a $N$ profile (red line) very similar to the fiducial model (blue line), but it contains a bell-shaped structure where there is the smooth change in density. This structure increases the integral $\int_{r_1}^{r_2} N /r \, dr$ compared to the fiducial model, thus, we expect a value of the asymptotic period spacing $\Delta P_\mathrm{a, Discont}$ lower than $\Delta P_\mathrm{a, Fiducial}$. Moreover, we expect that such bell-shaped structure in the $N$ profile becomes a glitch, because its width is much lower than the local wavelength of the waves \citep{Cunha2019,Cunha2024}. Therefore, we expect a sinusoidal behaviour of $\Delta P$ around $\Delta P_\mathrm{a, Smooth}$ with a decreasing amplitude at increasing period, and dips in the period spacing that are evenly spaced in period, with a periodicity compatible with the normalised buoyancy radius of the glitch. This structure is aligned identically with the glitch discussed in Section \ref{sec:translated_discont}, as we have placed the peak of the bell function at a location analogous to the $\delta$-distribution. To verify that the model behaves as predicted by theory, we calculate eigenmodes also outside the observable region of the spectrum.

In the bottom panel, the period spacing of the $\gamma$-modes as a function of the eigenperiods for this new model. The fiducial model is not displayed in the bottom panel, as it would be indistinguishable from $\Delta P_\mathrm{a, Smooth}$, which represents the asymptotic period spacing of the model with a smooth transition. As expected, there is a small difference between the two asymptotic period spacings, that is $\Delta P_\mathrm{a, Smooth} \approx \Delta P_\mathrm{a, Fiducial} - 0.86$ s. The period spacing shows a periodicity of approximately 51 minutes with a decreasing amplitude at increasing eigenperiods. This periodicity is compatible with the observed $\Delta n \approx 12.4$ and corresponds to a normalised buoyancy radius of 0.081, as described in equation \ref{eq:Tau_signal_Dk}. This indicates consistency with the normalised buoyancy radius of the bell-shaped structure in $N$, which peaks at $\Phi \approx 0.082$ and possesses a $\mathrm{FWHM} \approx 0.011$. We further reinforce the alignment of these findings with the characteristics of the bell-shaped structure by checking the relation between the bell-shaped structure and the observed glitch signature similarly to Section \ref{sec:glitch_boundary_cc}.
\begin{figure}[htbp]
\centering
\includegraphics[width=\columnwidth]{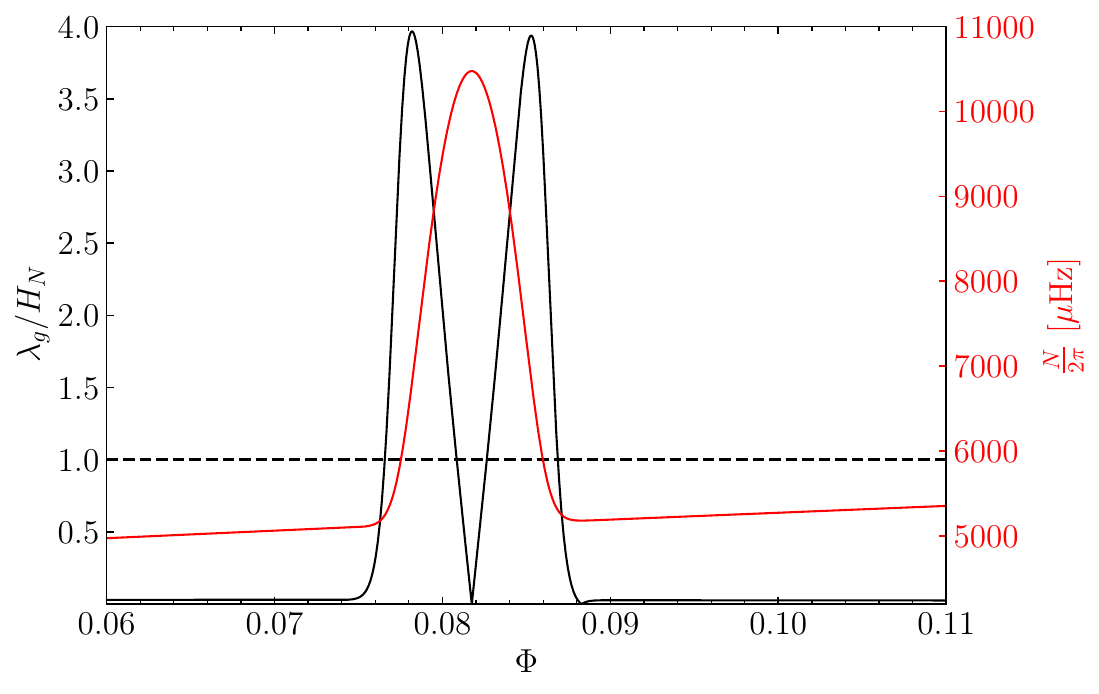} 
\caption{In black, we show the ratio of the local wavelength of g-modes ($\lambda_g$) near \numax to the Brunt-Väisälä length scale ($H_N$) as a function of the normalised buoyancy radius ($\Phi$) for a model characterised by a bell-shaped structure in the $N$ profile within the radiative core (refer to Section \ref{sec:smoothdiscont}). The dashed black line indicates the threshold where $\lambda_g/H_N = 1$. Additionally, we present the $N$ profile (red line) as a function of $\Phi$.}
\label{fig:HN_lambda_g_translated_smooth_100} 
\end{figure}
We find the presence of two structure variations where $H_N$ is significantly smaller than $\lambda_g$. These two glitches are located at $\Phi \approx 0.078$ and $\Phi \approx 0.085$ (see Figure \ref{fig:HN_lambda_g_translated_smooth_100}).
Therefore, contrary to the case with the $\delta$-distribution, we expect a glitch signature in the period spacing with visible beats attributable to these two adjacent sharp variations, which is clearly visible in the bottom panel of the figure. Furthermore, applying a Fourier transform to the period spacing data demonstrates that the two primary frequencies responsible for the beats correspond with the aforementioned peaks, collectively accounting for the observed periodicity. Finally, the small glitch explained in Section \ref{sec:fiducial_model} adds to the above glitch, generating relative fluctuations much lower than 1 \% around the main behaviour of the period spacing.
\begin{figure}[htbp]
\centering
\includegraphics[width=\linewidth]{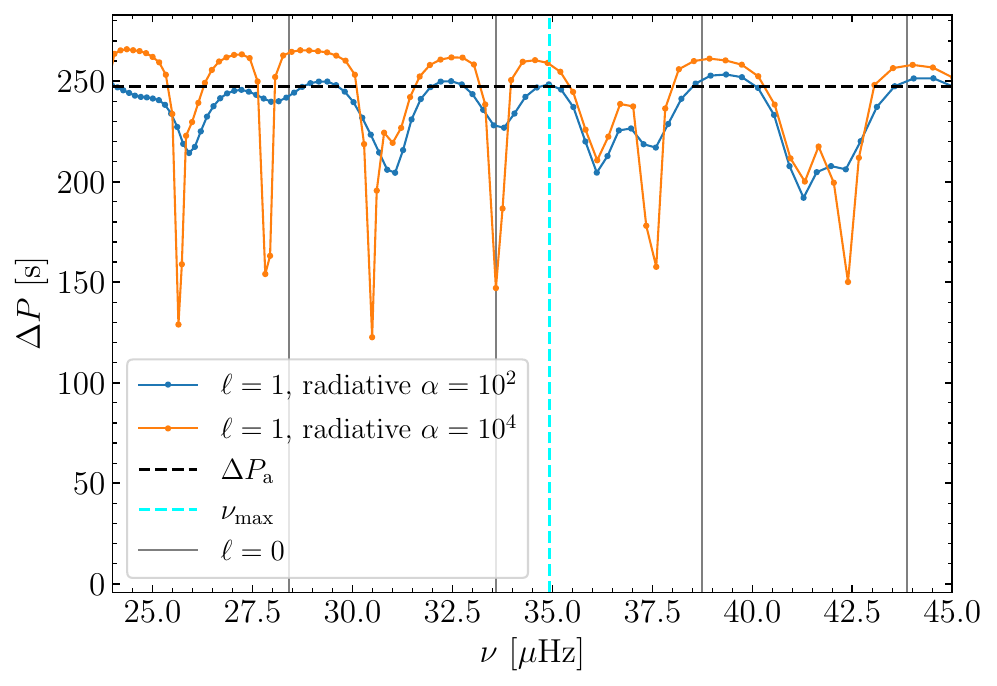} 
\caption{This figure displays a comparison of two period spacings of mixed dipole modes within the observable region of the spectrum, derived from different levels of smoothing applied in the radiative core. The cyan dashed line indicates \numax, while the black dashed line denotes the asymptotic $\Delta P$. Additionally, the grey lines correspond to the radial modes. The shown $\Delta P_\mathrm{a}$ and radial modes are representative of both models, as they are indistinguishable in this scale. Notably, we observe clear differences in their period spacings, even in areas where dipole modes are detectable.}
\label{fig:Smooth_translated_discont_PSD_comparison} 
\end{figure}

A more detailed characterisation of the detectability of the modes and the ability to differentiate between varying degrees of smoothing of the bell-shape structure is presented in Figure \ref{fig:Smooth_translated_discont_PSD_comparison}. This figure compares two period spacings of the mixed dipole modes within the observable region of the spectrum derived from different levels of smoothing (i.e. $\alpha = 10^2$ and $\alpha = 10^4$). Notably, we observe clear differences in their period spacings, even in areas where dipole modes are detectable. This finding highlights promising directions for interpreting glitch signatures in high-quality asteroseismic data, including the possibility to discern the sharpness of the density jump. However, one has to remember that not all of these minima show observable amplitudes and the measured $\Delta P$ is likely to differ from the true value if some modes are absent (as explained in Section \ref{sec:translated_discont}).


\subsection{Evolved CHeB stars}
\label{sec:diff_jump_discont_core}
All the barotropic models examined are derived from evolutionary models at the beginning of the CHeB phase (i.e. with $Y_c \approx 0.9$). However, as this evolutionary phase progresses, a growing density gradient is expected at the boundary between the convective and radiative core \citep[see e.g.][]{Kippenhahn2012}. Moreover, if the overshoot region exhibits radiative stratification, this density discontinuity is expected to occur within the radiative core, potentially giving rise to glitches within the observable region of the oscillation spectrum. In Section \ref{sec:obs_glitches}, we investigate the impact on the eigenfrequencies of convective overshooting beyond the boundary between the convective and radiative core in the context of more evolved stellar models. The overshoot region we model possesses radiative thermal stratification. To mitigate complications arising from the appearance of semiconvective layers, we evolve stellar models using \texttt{CLES} and \texttt{MESA} until reaching $Y_c \approx 0.6$. We then establish this new density gradient at the boundary as the new $\Lambda$, while maintaining the same total mass and radius\footnote{Additionally, we calibrated a barotropic model that incorporates the same physical parameters as the $Y_c \approx 0.6$ \texttt{CLES} model; however, this does not affect the primary conclusions regarding the observed glitches.} as the fiducial model of Section \ref{sec:fidmod}.
\begin{table}
\centering
\resizebox{\linewidth}{!}{%
\begin{threeparttable}
\centering
\caption{Comparison between the barotropic models described in Section \ref{sec:diff_jump_discont_core}.}
\label{tab:results_baro_old}
\begin{tabular}{@{}lllllll@{}}
\toprule
Location & $\alpha$ & HWHM [$H_P$] & $\Delta P_\mathrm{a}$ [s] & $\Delta n_\mathrm{glitch}$ & $P_\mathrm{glitch}$ [min]& $\Phi(r_\mathrm{glitch})$ \\ \midrule
Boundary &  $\infty$ & 0  &  258.67 & $\infty$ & $\infty$& 0 \\ \midrule
Boundary &  $10^3$  & 0.0012 &  255.7 & 84.6 & 360.7 & 0.0118 \\ \midrule
Boundary &  $10^2$  & 0.012  &  250.27 & 27.14 & 113.2 & 0.037 \\ \midrule
Boundary &  10   & 0.16 &  236.98 & 0 & 0 & - \\ \toprule
Radiative core & $\infty$ & 0 &  265.3 & 11.69 & 51.6 & 0.086 \\ \midrule
Radiative core &  50 & 0.03 &  250.92 & 8.50 & 35.5 & 0.117 \\ \midrule
Radiative core &  10 & 0.17 &  246.56 & 0 & 0 & - \\ \bottomrule
\end{tabular}
\tablefoot{The first column presents the location of a density discontinuity, which may occur at the boundary between the convective core and the radiative core or situated within the radiative core itself. The second and third columns refer to the width of the bell-shaped structures incorporated in the models. The fourth column presents the asymptotic period spacing, while the subsequent columns indicate the periodicity of the glitches that have been observed (if any) along with their inferred locations.
}
\end{threeparttable}
}
\end{table}
The approach for modelling the additional structural glitches follows the methodologies outlined in Sections \ref{sec:glitch_boundary_cc} and \ref{sec:glitch_translated_discont}.

\subsubsection{Detectable structural glitches}
\label{sec:obs_glitches}
The main results concerning the observability of structural glitches across the seven models are summarised in Table \ref{tab:results_baro_old}. This table details the periodicity of the detected glitches, if present, along with their inferred locations. In all these cases the observed glitches align with the predictions made by equation \ref{eq:Tau_signal_Dk}, but a strict interpretation of this equation is not always useful in determing the observability of structural variations. For instance, a structural glitch situated precisely at the boundary between the convective and radiative core would still yield a glitch signature in the eigenfrequencies, albeit with an infinite periodicity. Therefore, even if a structural variation exists, it would remain undetectable. A similar condition arises for the case where $\alpha = 10$; however, in this case, no glitch signature is observable because the structural variation lacks sufficient sharpness. This scenario also pertains to the hydrogen-burning shell discussed in Section \ref{sec:fiducial_model}. Finally, when the parameter $\alpha$ assumes a finite value, beats emerge in the period spacing similarly to Section \ref{sec:Smooth_translated_discont}.

\begin{figure}[htbp]
\centering
\includegraphics[width=\columnwidth]{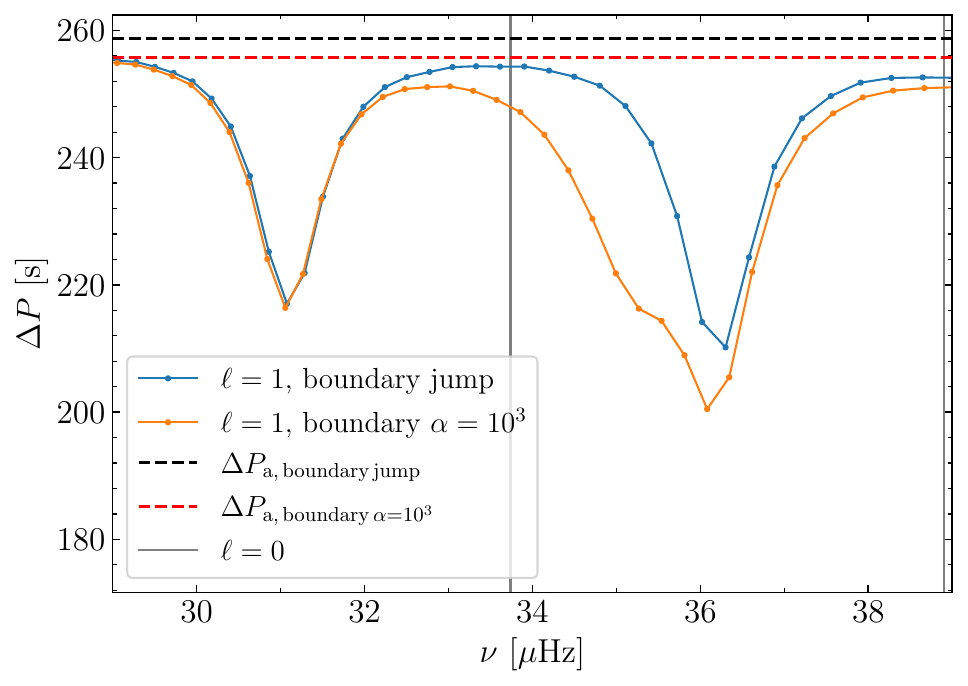} 
\caption{This figure displays a comparison of two period spacings of mixed dipole modes within the observable region of the spectrum, derived from different conditions at the core boundary. In particular, we show a model with a density jump (blue solid line), and a model with a smooth transition (orange solid line). The dashed lines denote the asymptotic period spacings for the model with a density jump (in black), and for the other model (in red). Furthermore, the grey lines correspond to the radial modes associated with both models, as they are indistinguishable in this scale. Notably, we observe clear differences in their period spacings, even in areas where dipole modes are detectable.}
\label{fig:Inertia_GAMMA_boundary_Old_CLES} 
\end{figure}
\begin{figure}[htbp]
\centering
\includegraphics[width=\columnwidth]{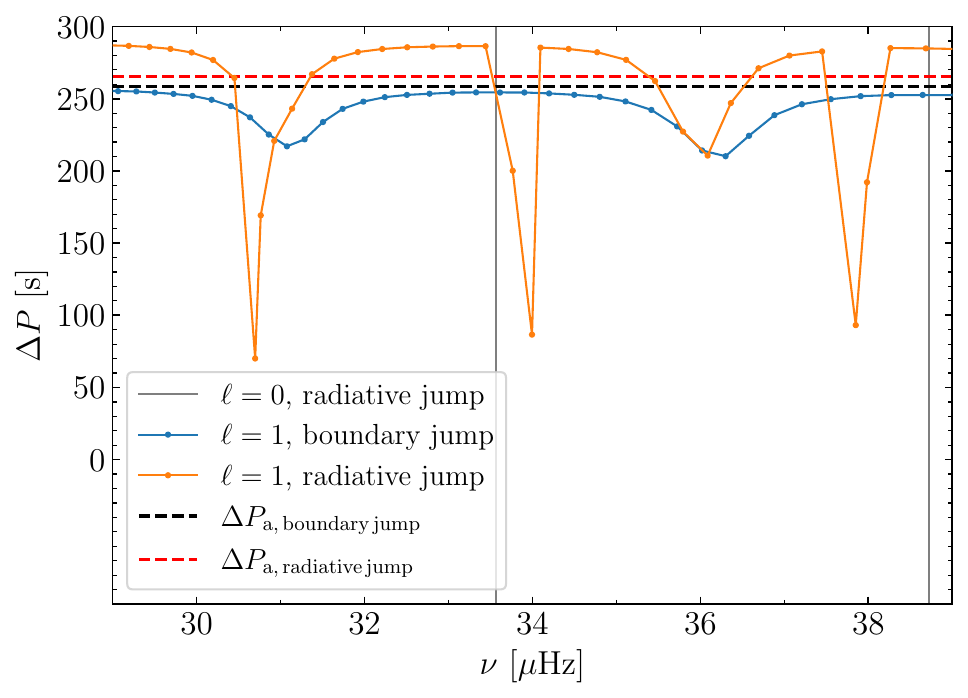} 
\caption{This figure displays a comparison of two period spacings of mixed dipole modes within the observable region of the spectrum. In particular, we show a model with a density jump located at the boundary between the convective and radiative core (blue solid line), and a model with a density jump located within the radiative core (orange solid line). The dashed lines denote the asymptotic period spacings for the model with a density jump (in black), and for the other model (in red). Furthermore, the grey lines correspond to the radial modes associated with the model with a density jump in the radiative core. Notably, we observe clear differences in their period spacings, even in areas where dipole modes are detectable.}
\label{fig:Inertia_GAMMA_radiative_Old_CLES} 
\end{figure}
The amplitude of the glitch signatures presented in this section is higher than that observed in Sections \ref{sec:glitch_boundary_cc} and \ref{sec:glitch_translated_discont} due to the higher $\Lambda$ density jump. Such behaviour is consistent with the theoretical expectations \citep[e.g.][]{Miglio2008,Cunha2015}. Consequently, it may be possible to observe these glitch signatures in the eigenfrequencies depending on the helium content in the centre of the CHeB star, as this content affects the density jump at the core boundary. Additionally, the observability of such glitches is influenced by the total radius of the CHeB star and/or its effective temperature, as both parameters alter \numax for a given total stellar mass. Ultimately, one has to consider that the observable region of the oscillation spectrum is limited, and not all modes within this range are detectable. Indeed, there is an inverse relationship between the inertia of a mode and its amplitude \citep[e.g.][]{Dupret2009}. Consequently, the inability to detect certain modes could hinder accurate identification of the location of the structural glitches and of the period spacing. Therefore, even when distinct differences among models are evident in the period spacing of the mixed dipole modes within the observable region of the spectrum (see Figures \ref{fig:Inertia_GAMMA_boundary_Old_CLES} and \ref{fig:Inertia_GAMMA_radiative_Old_CLES}), interference from glitch signatures with p-like dipole modes may further reduce the observable amplitude of the dipole modes, potentially resulting in their absence in the observed frequencies. These collective insights underline the complexities encountered in detecting glitch signatures in actual data \citep[e.g.][]{Vrard2022}.

\section{Summary and conclusions}
\label{sec:conc}
In this paper, we conducted a theoretical analysis of how structural variations adjacent to the convective core and chemical composition gradients within the radiative core influence the period spacing of mixed dipole modes and $\gamma$-modes in low-mass CHeB stars. These variations are also expected to occur within semiconvective layers, within the hydrogen-burning shell, and at the base of the convective envelope. Additionally, in low-mass stars with a degenerate helium core, the transition from RGB to CHeB is characterised by a succession of off-centre helium flashes, which induce chemical composition gradients in the radiative core. To investigate the impact of density discontinuities and associated structural glitches on the period spacings of these oscillation modes, we developed semi-analytical models of low-mass CHeB stars. These models were calibrated using the advanced evolutionary codes \texttt{BaSTI-IAC}, \texttt{CLES}, and \texttt{MESA}. The distinct physical prescriptions of these codes allowed us to identify common relevant features for calibration while enabling control over the type of structural glitch introduced, all while maintaining a realistic representation of stellar interiors. We first established a fiducial model based on a solar mass CHeB star at solar metallicity with $Y_c\approx 0.9$, featuring a realistic g-cavity. Subsequently, we explored the effects of various discontinuities in density, non-differentiable points within the density profile, and bell-shaped glitches in the Brunt–Väisälä frequency on the adiabatic eigenfrequencies.
The key findings from our analysis are summarised as follows:
\begin{itemize}
  \item Jump discontinuities in the density profile result in distinct glitch signatures in the period spacings of mixed modes. These glitches exhibit periodic behaviours that are closely tied to the normalised buoyancy radius associated with the discontinuities. While certain trapped modes may be observable, their detectability is influenced by their inertia, with higher inertia modes exhibiting lower amplitudes. This may lead to discrepancies between measured and true values of $\Delta P$ due to possible missing modes in observations. Notably, our analysis indicates that the inferred period spacing from observational data is more likely to represent the maximum measured value rather than the actual asymptotic one. However, they are typically close to each other.
  \item The comparison between models featuring smooth transitions and those with discontinuities highlighted differences in the periodic behaviours. Specifically, this finding suggests that smooth transitions can have an impact comparable to that of jump discontinuities (i.e. be characterised as structural glitches) if the scale height of the transition is much shorter than the local wavelength of the waves under investigation, presenting opportunities to detect not only the position and amplitude of the glitches but also their sharpness.
  \item Our simulations of 4-year-long \kepler observations reinforced the necessity of incorporating realistic stellar interior models to predict oscillation frequencies accurately. The resulting PSDs and period spacing patterns closely resemble observed data, providing a promising avenue for verifying our theoretical models against real-world observations.
  \item Additionally, we investigated more evolved CHeB stars with $Y_c \approx 0.6$. Our results indicate that the observability of glitch signatures is influenced by the helium content at the centre of the star, its total radius and/or its effective temperature. We also found that not all modes within the observable region of the spectrum are detectable, and the presence of glitch signatures near p-like mixed modes can decrease their observable amplitudes, complicating the correct identification of glitch signatures in actual data.
\end{itemize}
This work establishes a solid foundation for future asteroseismic studies aimed at probing the internal structures of stars. Our models enable realistic predictions of how each sharp structural variation impacts the observed power spectral density. This alignment not only validates our theoretical approach but also suggests promising directions for interpreting glitches signatures in high-quality asteroseismic data, such as that obtained from the \kepler mission, as well as from future space missions like the upcoming PLATO mission \citep{Rauer2025} and the HAYDN project \citep{Miglio2021b}.

\begin{acknowledgements}
We are grateful to Margarida Cunha, Arlette Noels and Richard Scuflaire for useful discussions. MM, AM, and WEvR acknowledge support from the ERC Consolidator Grant funding scheme (project ASTEROCHRONOMETRY, \url{https://www.asterochronometry.eu}, G.A. n. 772293). GB acknowledges fundings from the Fonds National de la Recherche Scientifique (FNRS) as a postdoctoral researcher. LP acknowledges fundings from the FNRS as a PhD student. We thank the anonymous referee for the helpful comments that improved the quality of the manuscript.
\end{acknowledgements}

\section*{Data Availability}
The data underlying this article will be shared on reasonable request to the corresponding author.



\bibliographystyle{aa}
\bibliography{paperbaro} 

\begin{thebibliography}{65}
\expandafter\ifx\csname natexlab\endcsname\relax\def\natexlab#1{#1}\fi

\bibitem[{{Aerts} {et~al.}(2010){Aerts}, {Christensen-Dalsgaard}, \&
  {Kurtz}}]{Aerts2010}
{Aerts}, C., {Christensen-Dalsgaard}, J., \& {Kurtz}, D.~W. 2010,
  {Asteroseismology} (Springer Dordrecht)

\bibitem[{{Asplund} {et~al.}(2009){Asplund}, {Grevesse}, {Sauval}, \&
  {Scott}}]{Asplund2009}
{Asplund}, M., {Grevesse}, N., {Sauval}, A.~J., \& {Scott}, P. 2009, \araa, 47,
  481

\bibitem[{{Ball} {et~al.}(2018){Ball}, {Chaplin}, {Schofield}, {Miglio},
  {Bossini}, {Davies}, \& {Girardi}}]{Ball2018}
{Ball}, W.~H., {Chaplin}, W.~J., {Schofield}, M., {et~al.} 2018, \apjs, 239, 34

\bibitem[{{Bedding} {et~al.}(2011){Bedding}, {Mosser}, {Huber},
  {Montalb{\'a}n}, {Beck}, {Christensen-Dalsgaard}, {Elsworth}, {Garc{\'\i}a},
  {Miglio}, {Stello}, {White}, {De Ridder}, {Hekker}, {Aerts}, {Barban},
  {Belkacem}, {Broomhall}, {Brown}, {Buzasi}, {Carrier}, {Chaplin}, {di Mauro},
  {Dupret}, {Frandsen}, {Gilliland}, {Goupil}, {Jenkins}, {Kallinger},
  {Kawaler}, {Kjeldsen}, {Mathur}, {Noels}, {Silva Aguirre}, \&
  {Ventura}}]{Bedding2011}
{Bedding}, T.~R., {Mosser}, B., {Huber}, D., {et~al.} 2011, \nat, 471, 608

\bibitem[{{Bossini} {et~al.}(2015){Bossini}, {Miglio}, {Salaris},
  {Pietrinferni}, {Montalb{\'a}n}, {Bressan}, {Noels}, {Cassisi}, {Girardi}, \&
  {Marigo}}]{Bossini2015}
{Bossini}, D., {Miglio}, A., {Salaris}, M., {et~al.} 2015, \mnras, 453, 2290

\bibitem[{{Bossini} {et~al.}(2017){Bossini}, {Miglio}, {Salaris}, {Vrard},
  {Cassisi}, {Mosser}, {Montalb{\'a}n}, {Girardi}, {Noels}, {Bressan},
  {Pietrinferni}, \& {Tayar}}]{Bossini2017}
{Bossini}, D., {Miglio}, A., {Salaris}, M., {et~al.} 2017, \mnras, 469, 4718

\bibitem[{{Castellani} {et~al.}(1971{\natexlab{a}}){Castellani}, {Giannone}, \&
  {Renzini}}]{Castellani1971b}
{Castellani}, V., {Giannone}, P., \& {Renzini}, A. 1971{\natexlab{a}}, \apss,
  10, 355

\bibitem[{{Castellani} {et~al.}(1971{\natexlab{b}}){Castellani}, {Giannone}, \&
  {Renzini}}]{Castellani1971a}
{Castellani}, V., {Giannone}, P., \& {Renzini}, A. 1971{\natexlab{b}}, \apss,
  10, 340

\bibitem[{{Chitre} \& {Shaviv}(1967)}]{Chitre1967}
{Chitre}, S.~M. \& {Shaviv}, G. 1967, Proceedings of the National Academy of
  Science, 57, 573

\bibitem[{{Constantino} {et~al.}(2015){Constantino}, {Campbell},
  {Christensen-Dalsgaard}, {Lattanzio}, \& {Stello}}]{Constantino2015}
{Constantino}, T., {Campbell}, S.~W., {Christensen-Dalsgaard}, J., {Lattanzio},
  J.~C., \& {Stello}, D. 2015, \mnras, 452, 123

\bibitem[{{Cox} \& {Giuli}(1968)}]{Cox1968}
{Cox}, J.~P. \& {Giuli}, R.~T. 1968, {Principles of stellar structure}

\bibitem[{{Cunha}(2020)}]{Cunha2020}
{Cunha}, M.~S. 2020, in Astrophysics and Space Science Proceedings, Vol.~57,
  Dynamics of the Sun and Stars; Honoring the Life and Work of Michael J.
  Thompson, ed. M.~J.~P.~F.~G. {Monteiro}, R.~A. {Garc{\'\i}a},
  J.~{Christensen-Dalsgaard}, \& S.~W. {McIntosh}, 185--196

\bibitem[{{Cunha} {et~al.}(2019){Cunha}, {Avelino}, {Christensen-Dalsgaard},
  {Stello}, {Vrard}, {Jiang}, \& {Mosser}}]{Cunha2019}
{Cunha}, M.~S., {Avelino}, P.~P., {Christensen-Dalsgaard}, J., {et~al.} 2019,
  \mnras, 490, 909

\bibitem[{{Cunha} {et~al.}(2024){Cunha}, {Damasceno}, {Amaral}, {Falorca},
  {Christensen-Dalsgaard}, \& {Avelino}}]{Cunha2024}
{Cunha}, M.~S., {Damasceno}, Y.~C., {Amaral}, J., {et~al.} 2024, \aap, 687,
  A100

\bibitem[{{Cunha} {et~al.}(2015){Cunha}, {Stello}, {Avelino},
  {Christensen-Dalsgaard}, \& {Townsend}}]{Cunha2015}
{Cunha}, M.~S., {Stello}, D., {Avelino}, P.~P., {Christensen-Dalsgaard}, J., \&
  {Townsend}, R.~H.~D. 2015, \apj, 805, 127

\bibitem[{{Deheuvels} \& {Belkacem}(2018)}]{Deheuvels2018}
{Deheuvels}, S. \& {Belkacem}, K. 2018, \aap, 620, A43

\bibitem[{{Deheuvels} {et~al.}(2016){Deheuvels}, {Brand{\~a}o}, {Silva
  Aguirre}, {Ballot}, {Michel}, {Cunha}, {Lebreton}, \&
  {Appourchaux}}]{Deheuvels2016}
{Deheuvels}, S., {Brand{\~a}o}, I., {Silva Aguirre}, V., {et~al.} 2016, \aap,
  589, A93

\bibitem[{{Dupret} {et~al.}(2009){Dupret}, {Belkacem}, {Samadi}, {Montalban},
  {Moreira}, {Miglio}, {Godart}, {Ventura}, {Ludwig}, {Grigahc{\`e}ne},
  {Goupil}, {Noels}, \& {Caffau}}]{Dupret2009}
{Dupret}, M.~A., {Belkacem}, K., {Samadi}, R., {et~al.} 2009, \aap, 506, 57

\bibitem[{{{\'E}rgma}(1971)}]{Ergma1971}
{{\'E}rgma}, {\'E}. 1971, \sovast, 15, 51

\bibitem[{{Freytag} {et~al.}(1996){Freytag}, {Ludwig}, \&
  {Steffen}}]{Freytag1996}
{Freytag}, B., {Ludwig}, H.~G., \& {Steffen}, M. 1996, \aap, 313, 497

\bibitem[{{Gabriel} \& {Scuflaire}(1979)}]{Gabriel1979}
{Gabriel}, M. \& {Scuflaire}, R. 1979, \actaa, 29, 135

\bibitem[{{Goldstein} \& {Townsend}(2020)}]{Goldstein2020}
{Goldstein}, J. \& {Townsend}, R.~H.~D. 2020, \apj, 899, 116

\bibitem[{{Gough}(2007)}]{Gough2007}
{Gough}, D.~O. 2007, Astronomische Nachrichten, 328, 273

\bibitem[{{Grosjean} {et~al.}(2014){Grosjean}, {Dupret}, {Belkacem},
  {Montalban}, {Samadi}, \& {Mosser}}]{Grosjean2014}
{Grosjean}, M., {Dupret}, M.~A., {Belkacem}, K., {et~al.} 2014, \aap, 572, A11

\bibitem[{{Harpaz}(1984)}]{Harpaz1984}
{Harpaz}, A. 1984, \mnras, 210, 633

\bibitem[{{Hatta}(2023)}]{Hatta2023}
{Hatta}, Y. 2023, \apj, 950, 165

\bibitem[{{Herwig}(2000)}]{Herwig2000}
{Herwig}, F. 2000, \aap, 360, 952

\bibitem[{{Hidalgo} {et~al.}(2018){Hidalgo}, {Pietrinferni}, {Cassisi},
  {Salaris}, {Mucciarelli}, {Savino}, {Aparicio}, {Silva Aguirre}, \&
  {Verma}}]{Hidalgo2018}
{Hidalgo}, S.~L., {Pietrinferni}, A., {Cassisi}, S., {et~al.} 2018, \apj, 856,
  125

\bibitem[{{Khan} {et~al.}(2018){Khan}, {Hall}, {Miglio}, {Davies}, {Mosser},
  {Girardi}, \& {Montalb{\'a}n}}]{Khan2018}
{Khan}, S., {Hall}, O.~J., {Miglio}, A., {et~al.} 2018, \apj, 859, 156

\bibitem[{{Kippenhahn} {et~al.}(2012){Kippenhahn}, {Weigert}, \&
  {Weiss}}]{Kippenhahn2012}
{Kippenhahn}, R., {Weigert}, A., \& {Weiss}, A. 2012, {Stellar Structure and
  Evolution} (Springer Berlin, Heidelberg)

\bibitem[{{Ledoux}(1947)}]{Ledoux1947}
{Ledoux}, P. 1947, \apj, 105, 305

\bibitem[{{Maeder}(1975)}]{Maeder1975}
{Maeder}, A. 1975, \aap, 40, 303

\bibitem[{{Matteuzzi} {et~al.}(2024){Matteuzzi}, {Hendriks}, {Izzard},
  {Miglio}, {Brogaard}, {Montalb{\'a}n}, {Tailo}, \& {Mazzi}}]{Matteuzzi2024}
{Matteuzzi}, M., {Hendriks}, D., {Izzard}, R.~G., {et~al.} 2024, \aap, 691, A17

\bibitem[{{Matteuzzi} {et~al.}(2023){Matteuzzi}, {Montalb{\'a}n}, {Miglio},
  {Vrard}, {Casali}, {Stokholm}, {Tailo}, {Ball}, {van Rossem}, \&
  {Valentini}}]{Matteuzzi2023}
{Matteuzzi}, M., {Montalb{\'a}n}, J., {Miglio}, A., {et~al.} 2023, \aap, 671,
  A53

\bibitem[{{McDermott}(1990)}]{McDermott1990}
{McDermott}, P.~N. 1990, \mnras, 245, 508

\bibitem[{{Michaud} {et~al.}(1984){Michaud}, {Fontaine}, \&
  {Beaudet}}]{Michaud1984}
{Michaud}, G., {Fontaine}, G., \& {Beaudet}, G. 1984, \apj, 282, 206

\bibitem[{{Michaud} {et~al.}(2010){Michaud}, {Richer}, \&
  {Richard}}]{Michaud2010}
{Michaud}, G., {Richer}, J., \& {Richard}, O. 2010, \aap, 510, A104

\bibitem[{{Miglio} {et~al.}(2021){Miglio}, {Girardi}, {Grundahl}, {Mosser},
  {Bastian}, {Bragaglia}, {Brogaard}, {Buldgen}, {Chantereau}, {Chaplin},
  {Chiappini}, {Dupret}, {Eggenberger}, {Gieles}, {Izzard}, {Kawata}, {Karoff},
  {Lagarde}, {Mackereth}, {Magrin}, {Meynet}, {Michel}, {Montalb{\'a}n},
  {Nascimbeni}, {Noels}, {Piotto}, {Ragazzoni}, {Soszy{\'n}ski}, {Tolstoy},
  {Toonen}, {Triaud}, \& {Vincenzo}}]{Miglio2021b}
{Miglio}, A., {Girardi}, L., {Grundahl}, F., {et~al.} 2021, Experimental
  Astronomy, 51, 963

\bibitem[{{Miglio} {et~al.}(2008){Miglio}, {Montalb{\'a}n}, {Noels}, \&
  {Eggenberger}}]{Miglio2008}
{Miglio}, A., {Montalb{\'a}n}, J., {Noels}, A., \& {Eggenberger}, P. 2008,
  \mnras, 386, 1487

\bibitem[{{Montalb{\'a}n} {et~al.}(2013){Montalb{\'a}n}, {Miglio}, {Noels},
  {Dupret}, {Scuflaire}, \& {Ventura}}]{Montalban2013}
{Montalb{\'a}n}, J., {Miglio}, A., {Noels}, A., {et~al.} 2013, \apj, 766, 118

\bibitem[{{Montalb{\'a}n} {et~al.}(2010){Montalb{\'a}n}, {Miglio}, {Noels},
  {Scuflaire}, \& {Ventura}}]{Montalban2010}
{Montalb{\'a}n}, J., {Miglio}, A., {Noels}, A., {Scuflaire}, R., \& {Ventura},
  P. 2010, \apjl, 721, L182

\bibitem[{{Montgomery} {et~al.}(2003){Montgomery}, {Metcalfe}, \&
  {Winget}}]{Montgomery2003}
{Montgomery}, M.~H., {Metcalfe}, T.~S., \& {Winget}, D.~E. 2003, \mnras, 344,
  657

\bibitem[{{Mosser} {et~al.}(2011){Mosser}, {Belkacem}, {Goupil}, {Michel},
  {Elsworth}, {Barban}, {Kallinger}, {Hekker}, {De Ridder}, {Samadi}, {Baudin},
  {Pinheiro}, {Auvergne}, {Baglin}, \& {Catala}}]{Mosser2011a}
{Mosser}, B., {Belkacem}, K., {Goupil}, M.~J., {et~al.} 2011, \aap, 525, L9

\bibitem[{{Mosser} {et~al.}(2024){Mosser}, {Dr{\'e}au}, {Pin{\c{c}}on},
  {Deheuvels}, {Belkacem}, {Lebreton}, {Goupil}, \& {Michel}}]{Mosser2024}
{Mosser}, B., {Dr{\'e}au}, G., {Pin{\c{c}}on}, C., {et~al.} 2024, \aap, 681,
  L20

\bibitem[{{Mosser} {et~al.}(2018){Mosser}, {Gehan}, {Belkacem}, {Samadi},
  {Michel}, \& {Goupil}}]{Mosser2018}
{Mosser}, B., {Gehan}, C., {Belkacem}, K., {et~al.} 2018, \aap, 618, A109

\bibitem[{{Mosser} {et~al.}(2012{\natexlab{a}}){Mosser}, {Goupil}, {Belkacem},
  {Marques}, {Beck}, {Bloemen}, {De Ridder}, {Barban}, {Deheuvels}, {Elsworth},
  {Hekker}, {Kallinger}, {Ouazzani}, {Pinsonneault}, {Samadi}, {Stello},
  {Garc{\'\i}a}, {Klaus}, {Li}, {Mathur}, \& {Morris}}]{Mosser2012c}
{Mosser}, B., {Goupil}, M.~J., {Belkacem}, K., {et~al.} 2012{\natexlab{a}},
  \aap, 548, A10

\bibitem[{{Mosser} {et~al.}(2012{\natexlab{b}}){Mosser}, {Goupil}, {Belkacem},
  {Michel}, {Stello}, {Marques}, {Elsworth}, {Barban}, {Beck}, {Bedding}, {De
  Ridder}, {Garc{\'\i}a}, {Hekker}, {Kallinger}, {Samadi}, {Stumpe}, {Barclay},
  \& {Burke}}]{Mosser2012b}
{Mosser}, B., {Goupil}, M.~J., {Belkacem}, K., {et~al.} 2012{\natexlab{b}},
  \aap, 540, A143

\bibitem[{{Mosser} {et~al.}(2017){Mosser}, {Pin{\c{c}}on}, {Belkacem},
  {Takata}, \& {Vrard}}]{Mosser2017}
{Mosser}, B., {Pin{\c{c}}on}, C., {Belkacem}, K., {Takata}, M., \& {Vrard}, M.
  2017, \aap, 600, A1

\bibitem[{{Mosser} {et~al.}(2015){Mosser}, {Vrard}, {Belkacem}, {Deheuvels}, \&
  {Goupil}}]{Mosser2015}
{Mosser}, B., {Vrard}, M., {Belkacem}, K., {Deheuvels}, S., \& {Goupil}, M.~J.
  2015, \aap, 584, A50

\bibitem[{{Noll} {et~al.}(2024){Noll}, {Basu}, \& {Hekker}}]{Noll2024}
{Noll}, A., {Basu}, S., \& {Hekker}, S. 2024, \aap, 683, A189

\bibitem[{{Ong} \& {Basu}(2020)}]{Ong2020}
{Ong}, J.~M.~J. \& {Basu}, S. 2020, \apj, 898, 127

\bibitem[{{Paxton} {et~al.}(2011){Paxton}, {Bildsten}, {Dotter}, {Herwig},
  {Lesaffre}, \& {Timmes}}]{Paxton2011}
{Paxton}, B., {Bildsten}, L., {Dotter}, A., {et~al.} 2011, \apjs, 192, 3

\bibitem[{{Paxton} {et~al.}(2019){Paxton}, {Smolec}, {Schwab}, {Gautschy},
  {Bildsten}, {Cantiello}, {Dotter}, {Farmer}, {Goldberg}, {Jermyn}, {Kanbur},
  {Marchant}, {Thoul}, {Townsend}, {Wolf}, {Zhang}, \& {Timmes}}]{Paxton2019}
{Paxton}, B., {Smolec}, R., {Schwab}, J., {et~al.} 2019, \apjs, 243, 10

\bibitem[{{Rauer} {et~al.}(2025){Rauer}, {Aerts}, {Cabrera}, {Deleuil},
  {Erikson}, {Gizon}, {Goupil}, {Heras}, {Walloschek}, {Lorenzo-Alvarez},
  {Marliani}, {Martin-Garcia}, {Mas-Hesse}, {O'Rourke}, {Osborn}, {Pagano},
  {Piotto}, {Pollacco}, {Ragazzoni}, {Ramsay}, {Udry}, {Appourchaux}, {Benz},
  {Brandeker}, {G{\"u}del}, {Janot-Pacheco}, {Kabath}, {Kjeldsen}, {Min},
  {Santos}, {Smith}, {Suarez}, {Werner}, {Aboudan}, {Abreu}, {Acu{\~n}a},
  {Adams}, {Adibekyan}, {Affer}, {Agneray}, {Agnor}, {Aguirre B{\o}rsen-Koch},
  {Ahmed}, {Aigrain}, {Al-Bahlawan}, {Alcacera Gil}, {Alei}, {Alencar},
  {Alexander}, {Alfonso-Garz{\'o}n}, {Alibert}, {Allende Prieto}, {Almeida},
  {Alonso Sobrino}, {Altavilla}, {Althaus}, {Alvarez Trujillo}, {Amarsi},
  {Ammler-von Eiff}, {Am{\^o}res}, {Andrade}, {Antoniadis-Karnavas},
  {Ant{\'o}nio}, {Aparicio del Moral}, {Appolloni}, {Arena}, {Armstrong},
  {Aroca Aliaga}, {Asplund}, {Audenaert}, {Auricchio}, {Avelino}, {Baeke},
  {Bailli{\'e}}, {Balado}, {Ballber Balaguer{\'o}}, {Balestra}, {Ball},
  {Ballans}, {Ballot}, {Barban}, {Barbary}, {Barbieri}, {Barcel{\'o} Forteza},
  {Barker}, {Barklem}, {Barnes}, {Barrado Navascues}, {Barragan}, {Baruteau},
  {Basu}, {Baudin}, {Baumeister}, {Bayliss}, {Bazot}, {Beck}, {Belkacem},
  {Bellinger}, {Benatti}, {Benomar}, {B{\'e}rard}, {Bergemann}, {Bergomi},
  {Bernardo}, {Biazzo}, {Bignamini}, {Bigot}, {Billot}, {Binet}, {Biondi},
  {Biondi}, {Birch}, {Bitsch}, {Bluhm Ceballos}, {B{\'o}di}, {Bogn{\'a}r},
  {Boisse}, {Bolmont}, {Bonanno}, {Bonavita}, {Bonfanti}, {Bonfils}, {Bonito},
  {Bonomo}, {B{\"o}rner}, {Boro Saikia}, {Borreguero Mart{\'\i}n}, {Borsa},
  {Borsato}, {Bossini}, {Bouchy}, {Bou{\'e}}, {Boufleur}, {Boumier},
  {Bourrier}, {Bowman}, {Bozzo}, {Bradley}, {Bray}, {Bressan}, {Breton},
  {Brienza}, {Brito}, {Brogi}, {Brown}, {Brown}, {Brun}, {Bruno}, {Bruns},
  {Buchhave}, {Bugnet}, {Buldgen}, {Burgess}, {Busatta}, {Busso}, {Buzasi},
  {Caballero}, {Cabral}, {Cabrero Gomez}, {Calderone}, {Cameron}, {Cameron},
  {Campante}, {Campos Gestal}, {Canto Martins}, {Cara}, {Carone}, {Carrasco},
  {Casagrande}, {Casewell}, {Cassisi}, {Castellani}, {Castro}, {Catala},
  {Catal{\'a}n Fern{\'a}ndez}, {Catelan}, {Cegla}, {Cerruti}, {Cessa},
  {Chadid}, {Chaplin}, {Charpinet}, {Chiappini}, {Chiarucci}, {Chiavassa},
  {Chinellato}, {Chirulli}, {Christensen-Dalsgaard}, {Church}, {Claret},
  {Clarke}, {Claudi}, {Clermont}, {Coelho}, {Coelho}, {Cogato}, {Colom{\'e}},
  {Condamin}, {Conde Garc{\'\i}a}, \& {Conseil}}]{Rauer2025}
{Rauer}, H., {Aerts}, C., {Cabrera}, J., {et~al.} 2025, Experimental Astronomy,
  59, 26

\bibitem[{{Schwarzschild}(1958)}]{Schwarzschild1958}
{Schwarzschild}, M. 1958, {Structure and evolution of the stars.}

\bibitem[{{Scuflaire} {et~al.}(2008{\natexlab{a}}){Scuflaire}, {Montalb{\'a}n},
  {Th{\'e}ado}, {Bourge}, {Miglio}, {Godart}, {Thoul}, \&
  {Noels}}]{Scuflaire2008b}
{Scuflaire}, R., {Montalb{\'a}n}, J., {Th{\'e}ado}, S., {et~al.}
  2008{\natexlab{a}}, \apss, 316, 149

\bibitem[{{Scuflaire} {et~al.}(2008{\natexlab{b}}){Scuflaire}, {Th{\'e}ado},
  {Montalb{\'a}n}, {Miglio}, {Bourge}, {Godart}, {Thoul}, \&
  {Noels}}]{Scuflaire2008a}
{Scuflaire}, R., {Th{\'e}ado}, S., {Montalb{\'a}n}, J., {et~al.}
  2008{\natexlab{b}}, \apss, 316, 83

\bibitem[{{Straniero} {et~al.}(2003){Straniero}, {Dom{\'\i}nguez}, {Imbriani},
  \& {Piersanti}}]{Straniero2003}
{Straniero}, O., {Dom{\'\i}nguez}, I., {Imbriani}, G., \& {Piersanti}, L. 2003,
  \apj, 583, 878

\bibitem[{{Takata}(2016)}]{Takata2016a}
{Takata}, M. 2016, \pasj, 68, 109

\bibitem[{{Townsend} {et~al.}(2018){Townsend}, {Goldstein}, \&
  {Zweibel}}]{Townsend2018}
{Townsend}, R.~H.~D., {Goldstein}, J., \& {Zweibel}, E.~G. 2018, \mnras, 475,
  879

\bibitem[{{Townsend} \& {Teitler}(2013)}]{Townsend2013}
{Townsend}, R.~H.~D. \& {Teitler}, S.~A. 2013, \mnras, 435, 3406

\bibitem[{{Unno} {et~al.}(1989){Unno}, {Osaki}, {Ando}, {Saio}, \&
  {Shibahashi}}]{Unno1989}
{Unno}, W., {Osaki}, Y., {Ando}, H., {Saio}, H., \& {Shibahashi}, H. 1989,
  {Nonradial oscillations of stars}

\bibitem[{{Vrard} {et~al.}(2022){Vrard}, {Cunha}, {Bossini}, {Avelino},
  {Corsaro}, \& {Mosser}}]{Vrard2022}
{Vrard}, M., {Cunha}, M.~S., {Bossini}, D., {et~al.} 2022, Nature
  Communications, 13, 7553

\bibitem[{{Vrard} {et~al.}(2018){Vrard}, {Kallinger}, {Mosser}, {Barban},
  {Baudin}, {Belkacem}, \& {Cunha}}]{Vrard2018}
{Vrard}, M., {Kallinger}, T., {Mosser}, B., {et~al.} 2018, \aap, 616, A94

\bibitem[{{Vrard} {et~al.}(2016){Vrard}, {Mosser}, \& {Samadi}}]{Vrard2016}
{Vrard}, M., {Mosser}, B., \& {Samadi}, R. 2016, \aap, 588, A87

\end{thebibliography}




\begin{appendix}

\section{Taylor series solutions near the centre of barotropic stars}
\label{app:Taylor_exp}
In Section \ref{sec:diffeq} we discuss the set of differential equations we use to solve the internal structure of a barotropic star knowing the initial conditions. However, we cannot solve the differential equations numerically from the exact centre, because there the term $\psi/ \xi^2$ would lead the Runge-Kutta method to errors. To avoid this problem, we can begin the numerical evaluation of the equations near the centre using Taylor expansions. For simplicity, let us assume for now that $\gamma(\theta) = \gamma_c (\theta/\theta_c)^B$ near the centre of the star. We then obtain
\begin{equation}
\label{eq:Tayl_expan}
\begin{dcases}
\theta(\xi) = \theta_c - \frac{\theta_c^3 \xi^2}{6 \beta_c \gamma_c} + \frac{\theta_c^5 \xi^4}{360 \beta_c^2 \gamma_c^2} \left[ 13 - 5 (B + \gamma_c )  \right] + o(\xi^6) \\
\frac{d \theta(\xi)}{d \xi} = - \frac{\theta_c^3 \xi}{3 \beta_c \gamma_c} + \frac{\theta_c^5 \xi^3}{90 \beta_c^2 \gamma_c^2} \left[ 13 - 5 (B + \gamma_c )  \right] + o(\xi^5) \\
\psi(\xi) = \frac{\theta_c \xi^3}{3} - \frac{\theta_c^3 \xi^5}{30 \beta_c \gamma_c} + \frac{\theta_c^5 \xi^7}{2520 \beta_c^2 \gamma_c^2}  \left[ 13 - 5 (B + \gamma_c) \right] + o(\xi^9),
\end{dcases}
\end{equation}
with which we completely solve the numerical issue. Moreover, it can be inferred from equation \ref{eq:Tayl_expan} that, in close proximity to the centre of a barotropic star, $\gamma(r)$ can be regarded as approximately constant. Therefore, very near the centre, the star can be effectively modelled as a polytrope.

\section{Differential equations near the surface of barotropic stars}
\label{app:num_issues}
The set of differential equations we use in Section \ref{sec:diffeq} creates numerical issues at both the centre of the star (as discussed in Appendix \ref{app:Taylor_exp}) and at its surface.
The issue observed at the surface arises from the ratio $\theta^2 /\beta$, where both $\theta$ and $\beta$ approach zero.
Therefore, it is essential to formulate an alternative set of differential equations that avoids the issue associated with $\theta^2 /\beta$ in this region. To achieve this, we introduce a Lane-Emden like variable ($\omega$) defined as
\begin{equation}
\label{eq:omega_def}
\frac{\partial \omega(\theta)}{\partial\theta} := \frac{\beta(\theta)}{\theta^2}
\end{equation}
or, equivalently,
\begin{equation}
\label{eq:omega_def2}
\omega(\theta) := \omega_c + \int_{\theta_c}^{\theta} \frac{\beta(t)}{t^2} \, d t
\,\, .
\end{equation}
By substituting either equation \ref{eq:omega_def} (or equation \ref{eq:omega_def2}) along with equation \ref{eq:mass_density_diffeq} into the expression for $d \omega / d \xi$, one obtain
\begin{equation}
\label{eq:domega_dxi_}
\frac{d \omega}{d \xi} = - \frac{\psi(\xi)}{\gamma(\omega) \xi^2}
\,\, .
\end{equation}
The resulting general differential equation, which combines equation \ref{eq:mass_density_diffeq} with equation \ref{eq:domega_dxi_}, takes the form
\begin{equation}
\label{eq:Generalised_Lane_Emden_diffeq}
\frac{d^2 \omega}{d \xi^2} + \left( \frac{d \omega}{d \xi} \right)^2 \frac{1}{\gamma(\omega)} \frac{\partial \gamma (\omega)}{\partial \omega} + \frac{2}{\xi} \frac{d \omega}{d \xi} + \frac{\theta(\omega)}{\gamma(\omega)}= 0,
\end{equation}
where the new initial conditions are $\omega(\xi = 0) = \omega_c$ and $ \left. \frac{d \omega }{d \xi} \right|_{\xi = 0}= 0$.
This new equation overcomes the mentioned numerical issue, provided that $\gamma \neq 0$, because as $\theta$ approaches zero we do not have to deal with an indeterminate form.

Equation \ref{eq:Generalised_Lane_Emden_diffeq} simplifies when $\gamma(\omega)$ is a constant equal to $\gamma_c$, because we obtain the Lane-Emden-alike equation
\begin{equation}
\label{eq:Lane_Emden}
\frac{d^2 \omega}{d \xi^2} +  \frac{2}{\xi} \frac{d \omega}{d \xi}+ \frac{\theta(\omega)}{\gamma_c}= 0 ,
\end{equation} 
with
\begin{equation}
\beta(\theta) = \beta_c \left( \frac{\theta}{\theta_c} \right)^{\gamma_c},
\end{equation}
and
\begin{equation}
\label{eq:theta_omega_Lane_Emden}
\theta(\omega) = 
\begin{dcases}
\left[ \frac{(\gamma_c -1)\theta_c^{\gamma_c}}{\beta_c} \omega \right]^{\frac{1}{\gamma_c -1}} := \phi^{\frac{1}{\gamma_c -1}}, \quad \mathrm{for} \quad \gamma_c \neq 1 \\
\theta_c \exp{ \left( \frac{\theta_c}{\beta_c} \omega \right) }, \quad \mathrm{for} \quad \gamma_c = 1.
\end{dcases}
\end{equation}
From equation \ref{eq:theta_omega_Lane_Emden} one obtains that $\phi_c = \theta_c^{\gamma_c -1}$ for $\gamma_c \neq 1$, and that $\omega_c = 0$ for $\gamma_c = 1$.
Therefore, equation \ref{eq:Lane_Emden} can be divided into the two differential equations
\begin{equation}
\label{eq:Lane_Emden_final}
\begin{dcases}
\frac{d^2 \phi}{d \xi^2} +  \frac{2}{\xi} \frac{d \phi}{d \xi} + \frac{(\gamma_c -1)\theta_c^{\gamma_c}}{\beta_c \gamma_c} \phi^{\frac{1}{\gamma_c -1}}= 0, \quad \phi \in [\theta_c^{\gamma_c -1},0], \quad  \gamma_c \neq 1 \\
\frac{d^2 \omega}{d \xi^2} +  \frac{2}{\xi} \frac{d \omega}{d \xi} + \theta_c \exp{ \left( \frac{\theta_c}{\beta_c} \omega  \right) }  = 0, \quad \omega \in [0,-\infty), \quad \gamma_c = 1, 
\end{dcases}
\end{equation}
and from $(\phi, \frac{d \phi}{d \xi})$ or $(\omega, \frac{d \omega}{d \xi})$ we can a posteriori evaluate
\begin{equation}
\psi(\xi) =
\begin{dcases}
\frac{\beta_c \gamma_c}{(\gamma_c -1)\theta_c^{\gamma_c}} \left( -\xi^2 \frac{d \phi }{d \xi} \right), \quad  \gamma_c \neq 1 \\
- \xi^2 \frac{d \omega }{d \xi}, \quad  \gamma_c = 1,
\end{dcases}
\end{equation}
and
\begin{equation}
N^2(\xi) = 
\begin{dcases}
4 \pi G \rho_1  \frac{\beta_c}{\theta_c^{\gamma_c} \phi(\xi)}  \left( \frac{\gamma_c}{\gamma_c -1} \frac{d \phi}{d \xi} \right)^2 \left( \frac{1}{\gamma_c} - \frac{1}{\Gamma_1} \right), \quad  \gamma_c \neq 1 \\
4 \pi G \rho_1  \frac{\theta_c}{\beta_c}  \left( \frac{d \omega}{d \xi} \right)^2 \left(\frac{\Gamma_1 - 1}{\Gamma_1} \right), \quad  \gamma_c = 1.
\end{dcases}
\end{equation}

\section{Numerical solver verification}
\label{app:plummer}
We validate the numerical solver thanks to known analytical solutions to equation \ref{eq:mass_density_diffeq} when $\gamma(\rho) = \gamma_c$ is a constant. One solution is the Plummer sphere (i.e. $\gamma_c = 6/5$), which is 
\begin{equation}
\begin{dcases}
\theta(\xi) = \theta_c \left( 1 + \frac{\theta_c^2 \xi^2 }{18 \beta_c}\right)^{-\frac{5}{2}} \\
\beta (\theta) = \beta_c \left( \frac{\theta}{\theta_c}\right)^{\frac{6}{5}} = \beta_c \left( 1 + \frac{\theta_c^2 \xi^2 }{18 \beta_c}\right)^{-3}\\
\psi(\xi) = \frac{\theta_c \xi^3}{3} \left( 1 + \frac{\theta_c^2 \xi^2 }{18 \beta_c}\right)^{-\frac{3}{2}}.
\end{dcases}
\end{equation}
Another solution is for $\gamma_c = 2$, which is
\begin{equation}
\begin{dcases}
\theta(\xi) = \frac{\sqrt{2\beta_c}}{\xi} \sin{ \left( \frac{\theta_c \xi}{\sqrt{2 \beta_c}} \right) } \\
\beta (\theta) = \beta_c \left( \frac{\theta}{\theta_c}\right)^{2} = \frac{2 \beta_c^2}{\theta_c^2 \xi^2} \sin^2{ \left( \frac{\theta_c \xi}{\sqrt{2 \beta_c}} \right) } \\
\psi(\xi) = \frac{(2 \beta_c)^{\frac{3}{2}}}{\theta_c^2} \left[ \sin{ \left( \frac{\theta_c \xi}{\sqrt{2 \beta_c}} \right) }  - \left( \frac{\theta_c \xi}{\sqrt{2 \beta_c}} \right) \cos{ \left( \frac{\theta_c \xi}{\sqrt{2 \beta_c}} \right) }  \right].
\end{dcases}
\end{equation}

\section{Tests on artificial glitches in the fiducial model}
\label{app:artif_glitches_fiducial_model}
Upon analysing $\Delta P$ at frequencies lower than \numax, we identify a behaviour resembling a small amplitude oscillation centred in $\Delta P_\mathrm{a}$. The periodicity of this oscillation is approximately 11.6 minutes, while the corresponding periodicity in radial order is $\Delta n \approx 2.80$. These periodicities are consistent with expectations based on equation \ref{eq:Tau_signal_Dk}. However, it is important to note that the Brunt-Väisälä frequency at the normalised buoyancy radius associated with $\Delta n \approx 2.80$ is smooth, which implies that it should not affect the eigenfrequencies with a glitch signature \citep[e.g.][]{Cunha2015}. Notably, this periodicity is accompanied by an alias observed at $\Delta n \approx 1.555$. When this alias is taken into consideration, we find that the associated normalised buoyancy radius is in good agreement with the characteristics of the hydrogen-burning shell, which peaks at $\Phi \approx 0.641$ and has a full width at half maximum (FWHM) of 0.126. Nonetheless, it is crucial to remain aware that this periodicity may also arise from numerical inaccuracies in the computation of the adiabatic frequencies and/or from the presence of the aforementioned non-differentiable point.
\begin{figure}[htbp]
\centering
\includegraphics[width=\linewidth]{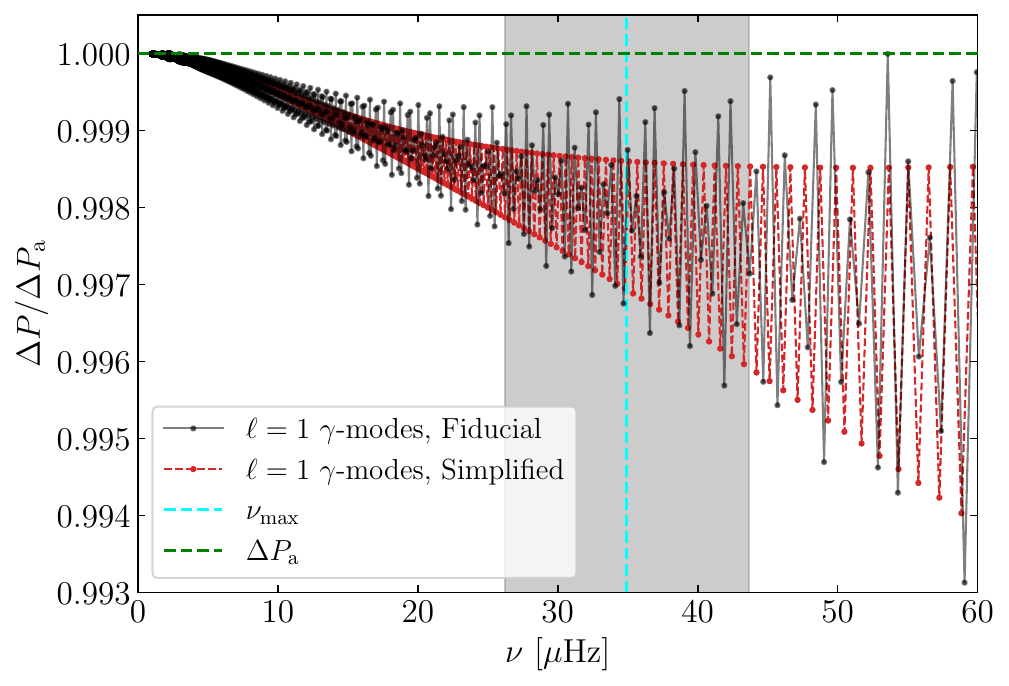} 
\caption[Comparison between normalised period spacings for the $\gamma$-modes as functions of frequency.]{Comparison between normalised period spacings for the $\gamma$-modes as functions of frequency. The grey area represents the observable region of the spectrum and the cyan line the \numax. It is clear that the fiducial model (in black) and the simplified model (in red) have similar properties of the g-cavity.}
\label{fig:fiducial_model_eigenfunction} 
\end{figure}
A way to solve this problem is by a comparison between the local wavelength of the eigenfunctions near the shell and the width of the glitch \citep[or the scale of variation of $N$, e.g.][]{Cunha2019}, because it shows whether there is a period trapping in the region and what the main source of the trapping is.
A simpler way to see buoyancy glitches, and conversely to solve the issue, is to measure the $\Delta P$ of the $\gamma$-modes. As can be seen from the top panel of the Figure \ref{fig:fiducial_model_brunt_inertia} and from Figure \ref{fig:fiducial_model_eigenfunction}, the small amplitude oscillations centred in the asymptotic period spacing $\Delta P_\mathrm{a}$ are more visible when the modulation caused by the coupling with the p-modes is absent.
While the aforementioned periodicity in radial order of these oscillations persists in the $\gamma$-modes, we must also account for a newly prominent glitch signature with periodicity $\Delta n \approx 2$, as evidenced by a Fourier transform applied to the period spacing of the oscillation modes. According to equation \ref{eq:Tau_signal_Dk}, we would expect a glitch structure at $\Phi(r_\mathrm{glitch}) \approx 0.5$. However, the $N$ function is smooth at this location, which means that it cannot generate a glitch signature in the eigenfrequencies. It is now apparent that the glitch at $\Delta n \approx 1.555$ aligns with the non-differentiable point, as at this specific location the scale of variation of $N$ is much lower than the local wavelength of the g-modes near \numax.
At this point, we can devise a simplified model that eliminates such non-differentiable point, albeit at the expense of a less realistic g-cavity. In particular, we employ the equations
\begin{equation}
\label{eq:simplified_model}
\begin{dcases}
\gamma(\theta) = A \exp{ \left[  - \frac{\left( \ln \theta - \ln \theta_P  \right)^2  }{2 \sigma^2}    \right]   }  + c , \quad \theta \le 1 \\\\
\ln \beta(\theta) = A \sqrt{\frac{\pi}{2}} \sigma\left[ \mathrm{erf} \left( \frac{\ln \theta - \ln \theta_P}{ \sqrt{2} \sigma} \right)  + \mathrm{erf} \left( \frac{ \ln \theta_P}{ \sqrt{2} \sigma} \right)  \right] + \\
+ c \ln \theta , \quad \theta \le 1
\end{dcases}
\end{equation}
instead of the equations \ref{eq:gamma_second_zone} and \ref{eq:ln_beta_second_zone}, with $c>0$ a constant chosen in order to have the correct total mass of the star. Figure \ref{fig:fiducial_model_eigenfunction} illustrates the normalised period spacings for the $\gamma$-modes from both the fiducial model (in black) and the simplified model.
The results demonstrate that the observed glitch signature with periodicity $\Delta n \approx 2$ remains intact, and that this oscillation appears to be at least compatible, if not predominant, with the glitch signature originating from the non-differentiable point. To further investigate this, we employed a finer grid and utilised different numerical solvers. Despite these modifications, the same glitches in the period spacing were consistently observed.
In particular, increasing the number of grid points by an order of magnitude results in a relative change in period spacing of the order of $10^{-2} \%$, while switching to a lower order numerical solver results in a relative change of the order of $10^{-5} \%$. Once the model has converged, these deviations from the mean period spacing do not decrease as the number of grid points increases.
Therefore, the main deviations from the asymptotic value could arise from numerical inaccuracies in the computation of eigenfrequencies, rather than from the presence of the non-differentiable point. Notably, the peak-to-peak relative difference in the period spacing of the modes is at most 0.2 \% (in the high-frequency regime), which is significantly lower than typical observational uncertainties \citep[e.g., $\approx 3$ s in the asymptotic period spacing of CHeB stars in the \kepler database,][]{Vrard2016,Vrard2022}. Consequently, this glitch does not compromise the interpretation of the observed power spectral distributions. Furthermore, it confirms that the other glitches, which will be addressed in sections \ref{sec:glitch_boundary_cc}, \ref{sec:glitch_translated_discont} and \ref{sec:diff_jump_discont_core}, can be examined without concern for interference from this particular glitch, as those will likely be much more significant.

\section{Non-differentiable continuous density function at the boundary between convective and radiative core}
\label{sec:no_discont}
\begin{figure}[htbp]
\centering
\begin{subfigure}[b]{\linewidth}
\centering
\includegraphics[width=\linewidth]{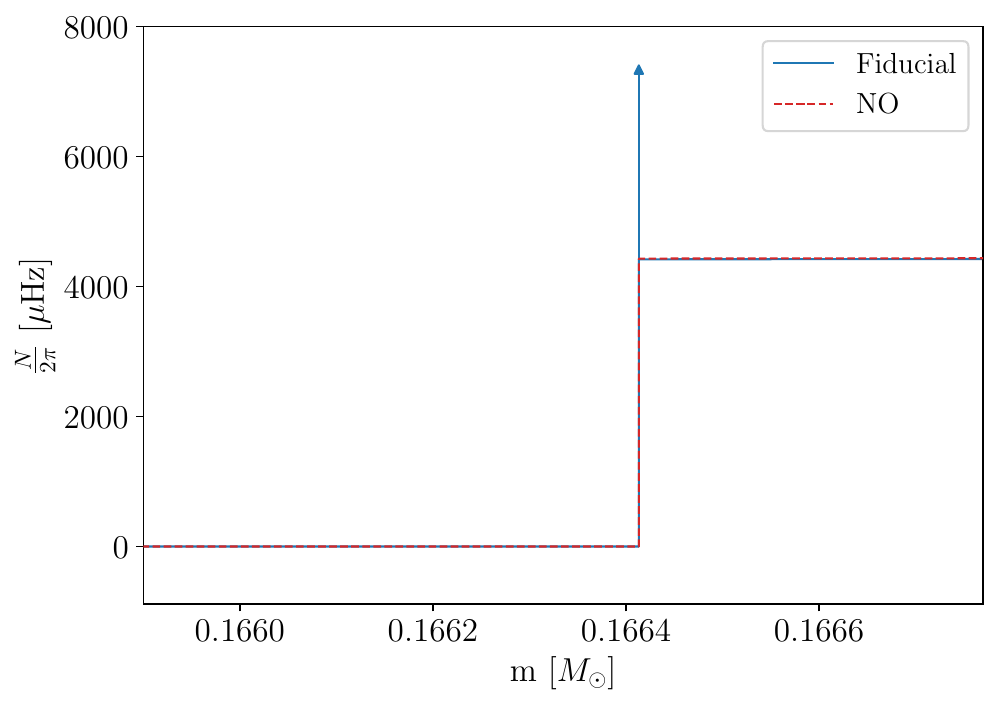} 
\end{subfigure}
\begin{subfigure}[b]{\linewidth}
\centering
\includegraphics[width=\linewidth]{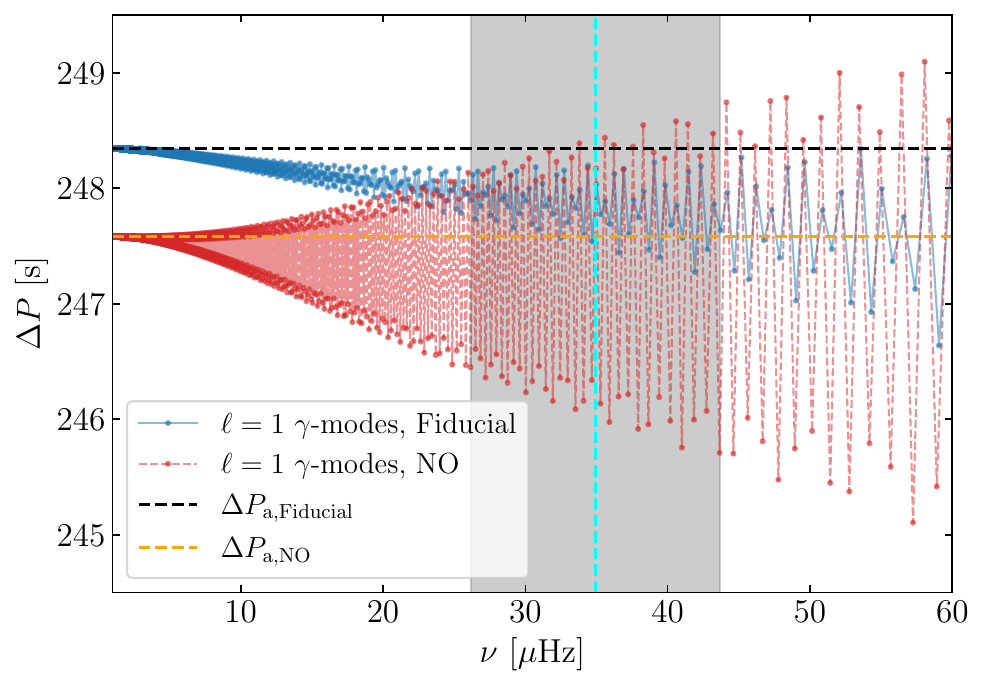} 
\end{subfigure}
\caption{Comparison of $N$ profiles as functions of internal mass (top panel) and period spacings of the $\gamma$-modes as functions of the eigenfrequencies (bottom panel). The blue line represents the fiducial model of Section \ref{sec:fidmod}, while the red model features a step-like structure in the $N$ profile at the boundary between the convective and radiative core instead of a $\delta$-distribution.}
\label{fig:no_discont} 
\end{figure}
In Figure \ref{fig:no_discont} we show in the top panel the $N$ profile as a function of the internal mass for a model similar to the fiducial barotropic model of Section \ref{sec:fidmod}, but with a continuous density function that is non-differentiable at the boundary between the convective and the radiative core. This has been done by choosing $\Lambda =1$ at the boundary. This new model has a $N$ profile (red line) very similar to the fiducial model (blue line), but it contains a jump discontinuity at the boundary instead of a $\delta$-distribution, and the bell-shaped peak related to the H-burning shell is located at a lower radius than in the fiducial model. Therefore, $\int_{r_1}^{r_2} N /r \, dr$ is higher in the new model and we expect a value of the asymptotic period spacing $\Delta P_\mathrm{a, NO}$ lower than $\Delta P_\mathrm{a, Fiducial}$.

In the bottom panel, the period spacings are presented as functions of the eigenfrequencies for both the fiducial model (in blue) and the new model (in red). Although the two asymptotic period spacings differ, the difference is minimal ($\approx 0.76$ s). Notably, the small glitch discussed in Section \ref{sec:fiducial_model} now shows an increased peak-to-peak relative difference in the period spacing of the modes. This could be related to the change of the $N$ profile at the boundary. However, it is important to note that the JWKB approximation used by \citet{Cunha2019,Cunha2024} cannot be applied in this context to infer the properties of the glitches. Despite this increased difference, the relative difference remains at most 1 \% in the observable region of the spectrum. As a result, it is very difficult to detect such a glitch even with the highest quality data available, especially when analysing mixed dipole modes.

\section{Modelling of the convective boundary mixing}
\label{sec:boundary_layer_MESA}
To address the convective boundary mixing problem, we employ phenomenological modelling with the same \texttt{MESA} tool applied in Section \ref{sec:fidmod}. For these simulations, we focus on a 1 \msol CHeB star with solar metallicity, employing identical undershooting prescriptions as described in Section \ref{sec:fidmod}. Specifically, we implement an exponential overshooting prescription \citep[e.g.][]{Maeder1975,Freytag1996,Herwig2000}, in which the overshoot region is defined by a temperature gradient $\nabla = \nabla_\mathrm{rad}$. 
The chosen prescription modifies the diffusive mixing coefficient $D_\mathrm{OV}$ near the boundary of the convective core such that
\begin{equation}
    D_\mathrm{OV} = D_0 \exp\left(\frac{-2 (r - r_0)}{0.035 H_{p,cc}}\right),
\end{equation}
where $H_{p,cc}$ represents the pressure scale height at the convective core boundary, $r$ is the radial coordinate, $D_0$ is the mixing coefficient at $r_0 = r_\mathrm{cc} - 0.005 H_{p,cc}$, and $r_\mathrm{cc}$ denotes the radius of the convective core boundary. Additionally, we implement a minimum mixing coefficient $D_\mathrm{min} = 1 \, \mathrm{cm}^2/\mathrm{s}$ in the internal structure of the star.

\end{appendix}

\end{document}